\documentclass[a4paper,fleqn]{cas-dc}
\let\printorcid\relax
\usepackage[numbers]{natbib}
\usepackage{hyperref}
\usepackage{lineno,hyperref}
\usepackage{graphics}
\usepackage{graphicx}
\usepackage{geometry}
\usepackage[svgnames]{xcolor}%
\usepackage{stfloats}
\usepackage{midfloat}
\usepackage[T1]{fontenc}
\usepackage{textcomp, amssymb}
\usepackage[utf8]{inputenc}
\usepackage{mathtools, cuted}
\usepackage{mathastext}
\usepackage{multicol}
\usepackage{caption}
\usepackage{adjustbox}

\begin{document}
\let\WriteBookmarks\relax
\def\floatpagepagefraction{1}
\def\textpagefraction{.001}
\shorttitle{}
\shortauthors{E. Lavrik et al.}

\title [mode = title]{Optical Inspection of the Silicon Micro-strip Sensors for the CBM Experiment employing 
%the Application of 
Artificial Intelligence  }                      
%\tnotemark[1,2]

%\tnotetext[1]{This document is the results of the research
%   project funded by the National Science Foundation.}

%\tnotetext[2]{The second title footnote which is a longer text matter
%   to fill through the whole text width and overflow into
%   another line in the footnotes area of the first page.}

\author[1]{E. Lavrik}[]
\cormark[1]
%\fnmark[1]
\ead{E.Lavrik@gsi.de}
%\ead[url]{}

%\credit{}

\address[1]{Facility for Antiproton and Ion Research in Europe, Planckstra{\ss}e 1, 64291
Darmstadt, Germany}

\address[2]{GSI Helmholtzzentrum f$\ddot{u}$r Schwerionenforschung GmbH, Planckstra{\ss}e 1, 64291
Darmstadt, Germany}

\address[3]{Helmholtz Forschungsakademie Hessen f$\ddot{u}$r FAIR (HFHF),  Max-von-Laue-Stra{\ss}e 12, 60438 Frankfurt am Main, Germany}

\author[2, 3, 4]{M. Shiroya}[]
\cormark[1]
%\fnmark[1]
\ead{M.Shiroya@gsi.de}

\author%
[2, 5]{H.R. Schmidt}[]

\author[2, 3, 4]{A. Toia}[]
%\fnmark[2]
%\ead{}

\address[4]{Institut f$\ddot{u}$r Kernphysik, Goethe Universit$\ddot{a}$t Frankfurt, Max-von-Laue-Stra{\ss}e 1,
60438, Frankfurt am Main, Germany}

\author%
[2]{J.M. Heuser}
%\cormark[2]
%\fnmark[1]
%\ead{}
%\ead[URL]{}

\address[5]{Physikalisches Institut,  Universit$\ddot{a}$t T$\ddot{u}$bingen, Auf der Morgenstelle 14,
72076, T$\ddot{u}$bingen, Germany}

\cortext[cor1]{Corresponding authors}
%\cortext[cor2]{Corresponding author}

\begin{abstract}
Optical inspection of 1191 silicon micro-strip sensors was performed using a custom made optical inspection setup, employing a machine-learning based approach for the defect analysis and subsequent quality assurance. Furthermore, metrological control of the sensor's surface was performed. In this manuscript, we present the analysis of various sensor surface defects. Among these are implant breaks, p-stop breaks, aluminium strip opens, aluminium strip shorts, surface scratches, double metallization layer defects, passivation layer defects, bias resistor defects as well as dust particle identification. The defect detection was done using the application of Convolutional Deep Neural Networks (CDNNs). From this, defective strips and defect clusters were identified, as well as a 2D map of the defects using their geometrical positions on the sensor was performed. Based on the total number of defects found on the sensor's surface, a method for the estimation of sensor's overall quality grade and quality score was proposed.

\end{abstract}


\maketitle
%\linenumbers
\section{Introduction}
Silicon sensors play a central role in modern High Energy Particle Physics experiments for tracking and momentum determination of the charged particles due to their fast response, high radiation tolerance, and good spatial resolution. At the same time, the complexity of the experiments and limited access during the detector operation puts stringent requirements onto the Quality Assurance (QA) of the silicon sensors. We already have reported on the devices and methods to perform automated sensor defect searches \cite{Lavrik}. In this manuscript, we report on the application of these procedures to a large sample of the sensors manufactured by Hamamatsu Photonics for the Compressed Baryonic Matter (CBM) experiment, comprising 1191 pieces overall.
CBM, a fixed-target heavy-ion physics experiment at the FAIR facility in Darmstadt, Germany, is designed to reconstruct heavy ion collisions at very high collision rates up to 10 MHz. Its main tracking detector, the Silicon Tracking System (STS), is currently under construction, comprising 876 detector modules made out of double-sided silicon micro-strip sensors of various sizes: 6.2 $\times$ 2.2 cm$^{2}$, 6.2 $\times$ 4.2 cm$^{2}$, 6.2 $\times$ 6.2 cm$^{2}$, 6.2 $\times$ 12.4 cm$^{2}$, whose further specifications are detailed in \cite{STS, cbmcollab}. The compact design of the tracking system will not allow for extensive maintenance and repair at the sensor level after the module assembly and installation inside the dipole magnet. Thus, as a first step in building the silicon tracker, a rigorous sensor QA is mandatory.

Hamamatsu Photonics, the producer of the silicon micro-strip sensors, could carry out a certain set of basic quality checks, i.e., detection of pin-holes and aluminum strip defects, and could guarantee the total amount of defective strips below 2$\%$.

However, other sensor defects can not be detected by the manufacturer's quality control, notably these defects can only be detected with an optical scanning of the sensor's surface using a microscope or with a set of specially designed electrical quality checks \cite{Panasenko}. 
Among these are implant strip breaks, double metallization layer defects, p-stop implantation breaks, bias resistor defects, etc. 
Apart from these defects, active or improper handling of the sensors might also affect their surface integrity, e.g. introduce surface scratches, lead to deposition of dust grains, etc.
Therefore, it is important to get a full view of all these defect types before assembling silicon sensor modules.
In the past, the ATLAS and CMS experiments at Large Hadron Collider (LHC) \cite{ATLAS, CMS} have reported 
on their quality control procedures involving electrical quality inspection along with visual inspection for their respective large silicon trackers \cite{Baumann, Carter, Krammer}.

\begin{figure*}
\centering
%\begin{center}
\includegraphics[height=8.1cm, width = 7.3cm]{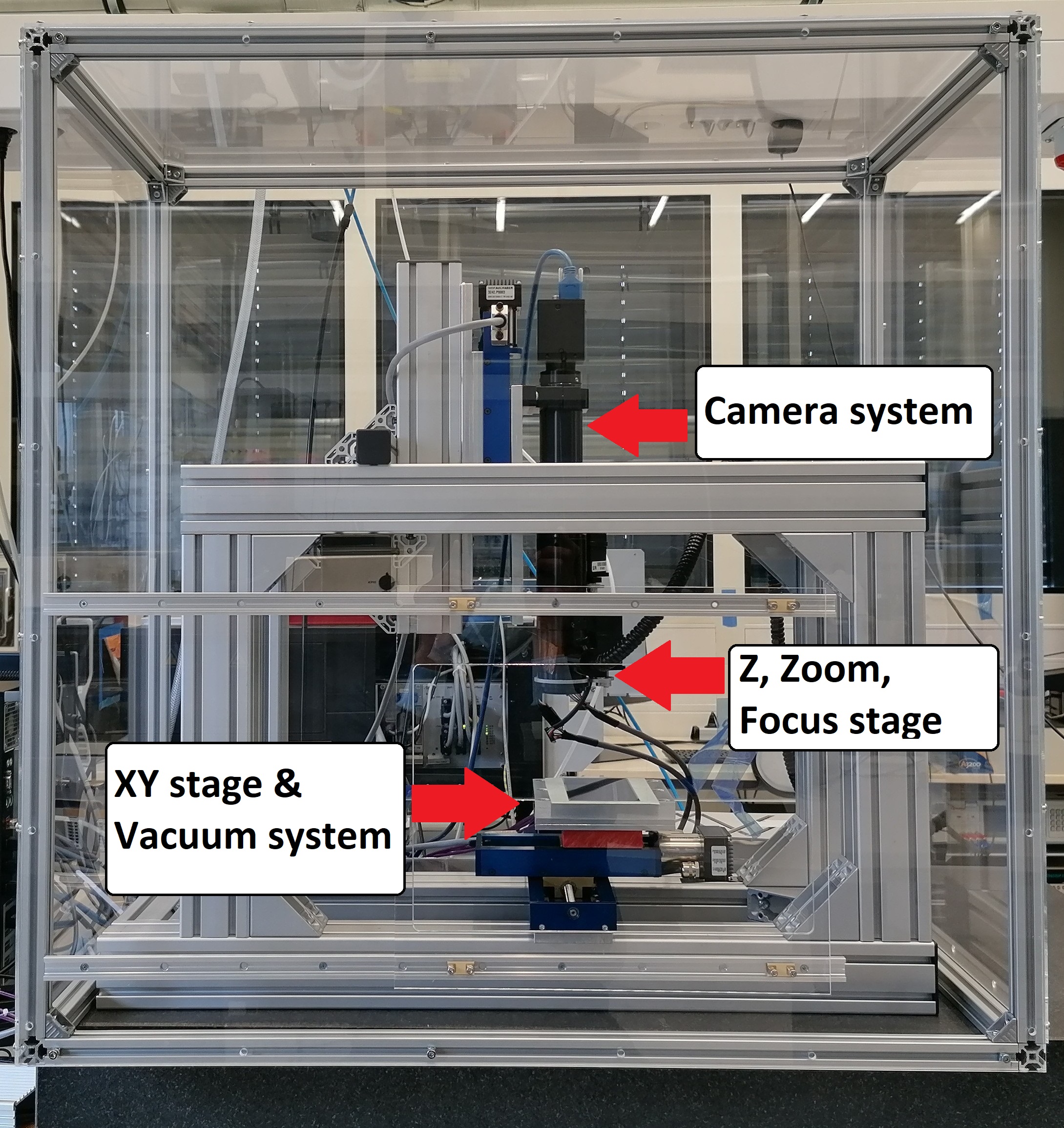}
\caption{Optical inspection setup located in a clean room of the Detector Laboratory department at GSI.}
%\end{center}
\label{fig:Setup}
\end{figure*}

\begin{figure*}
\centering
%\begin{center}
\includegraphics[height=4.2cm, width = 4.2cm]{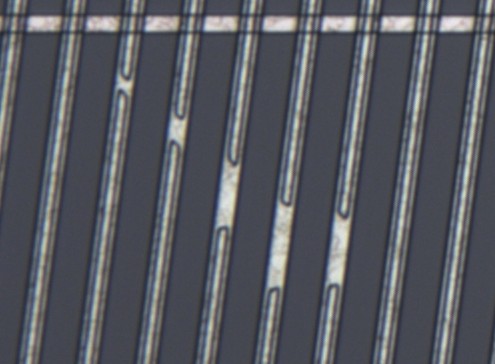}
\includegraphics[height=4.2cm, width = 4.2cm]{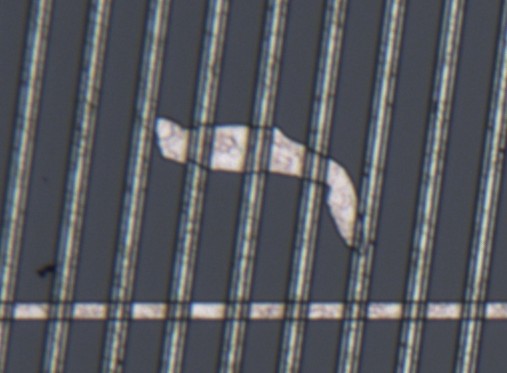}
\includegraphics[height=4.2cm, width = 4.2cm]{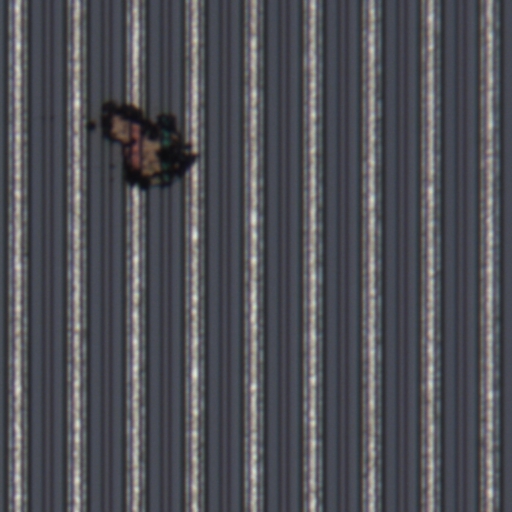}
\includegraphics[height=4.2cm, width = 4.2cm]{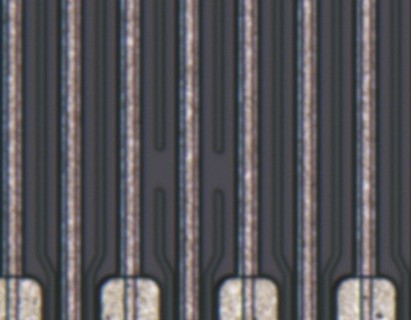}\\[1mm]
\includegraphics[height=4.2cm, width = 4.2cm]{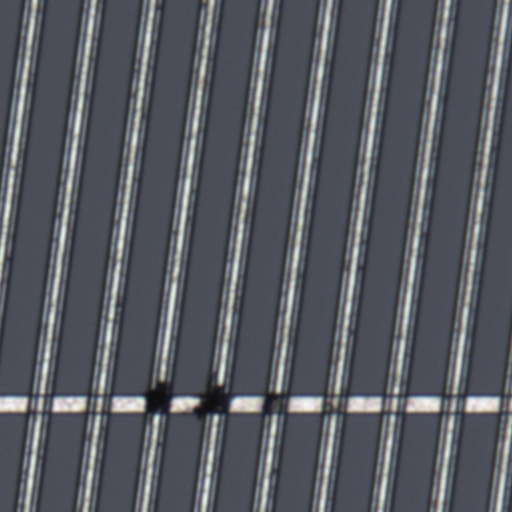}
\includegraphics[height=4.2cm, width = 4.2cm]{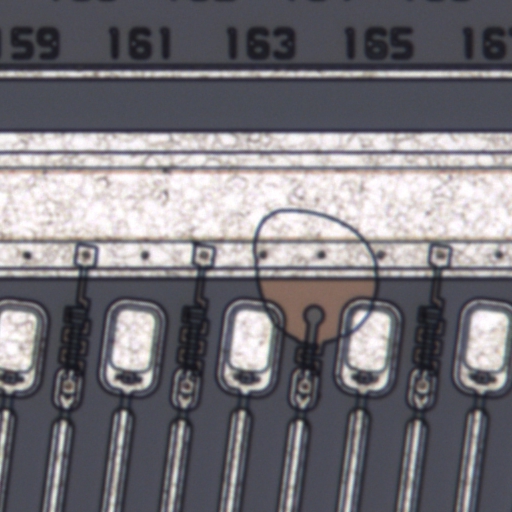}
\includegraphics[height=4.2cm, width = 4.2cm]{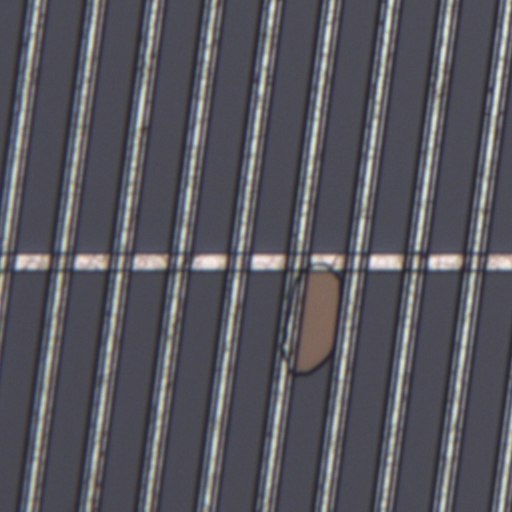}
\includegraphics[height=4.2cm, width = 4.2cm]{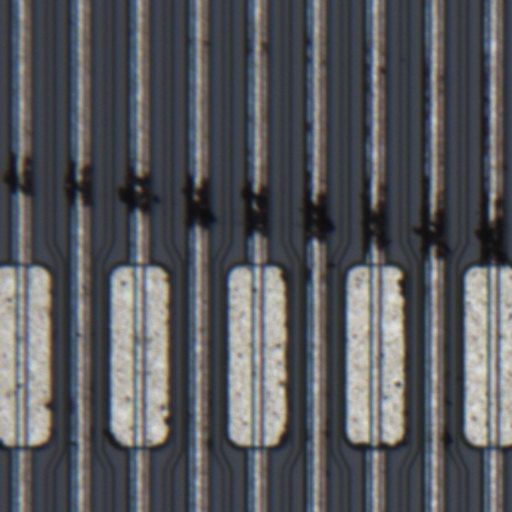}
\caption{Source images of surface defects taken from the optical inspection setup. Top panel: Implant break, Aluminium short, Aluminium open, P-stop break (From left to right) Bottom panel: Double metal line break, Resistor fault, Passivation open, Scratch (From left to right).}
%\end{center}
\label{fig:surface_defects}
\end{figure*} 

\begin{figure*}
\centering
%\begin{center}
\includegraphics[height=4.2cm, width = 4.2cm]{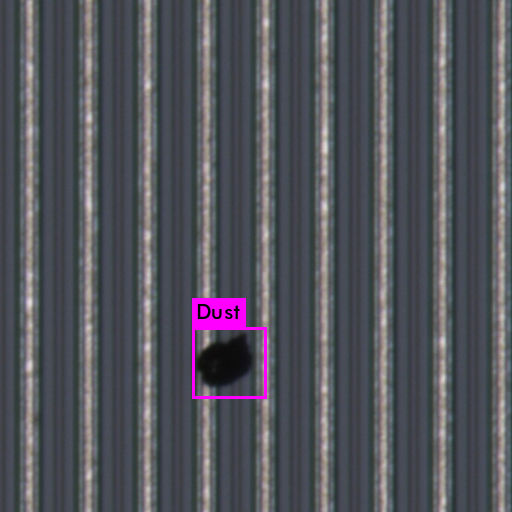}
\includegraphics[height=4.2cm, width = 4.2cm]{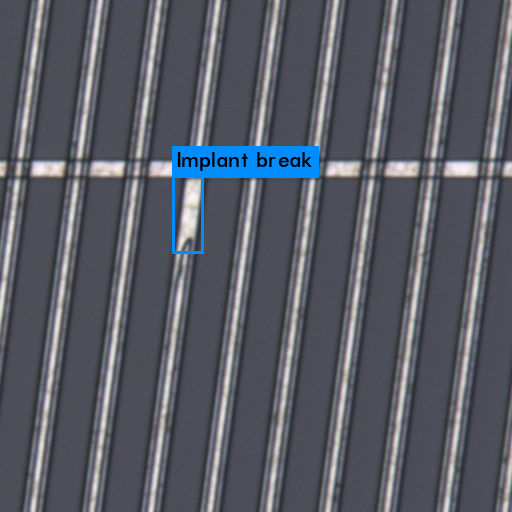}
\includegraphics[height=4.2cm, width = 4.2cm]{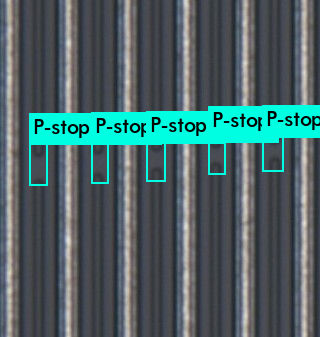}
\includegraphics[height=4.2cm, width = 4.2cm]{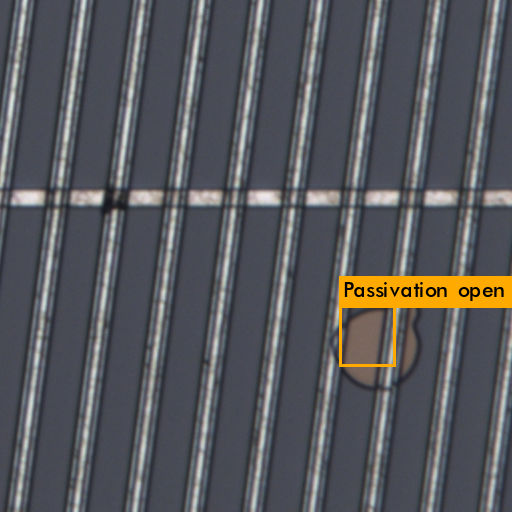}\\[1mm]
\includegraphics[height=4.2cm, width = 4.2cm]{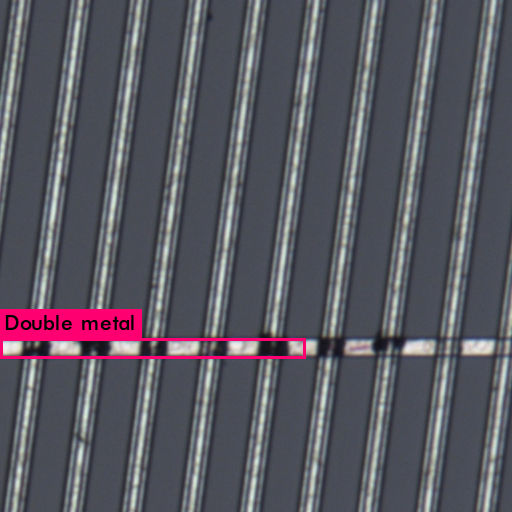}
\includegraphics[height=4.2cm, width = 4.2cm]{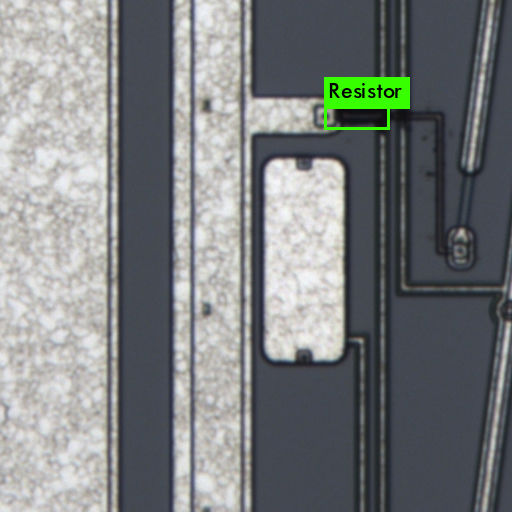} 
\includegraphics[height=4.2cm, width = 4.2cm]{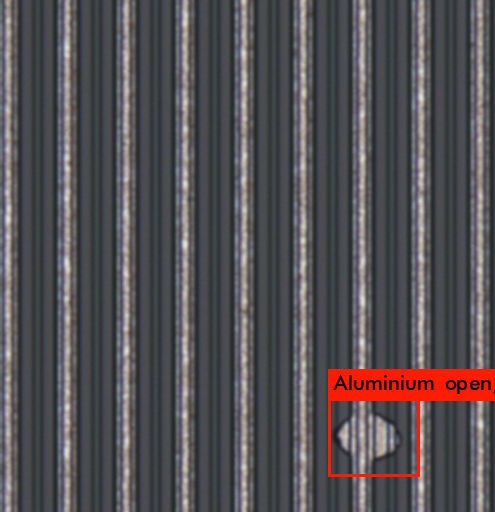}
\includegraphics[height=4.2cm, width = 4.2cm]{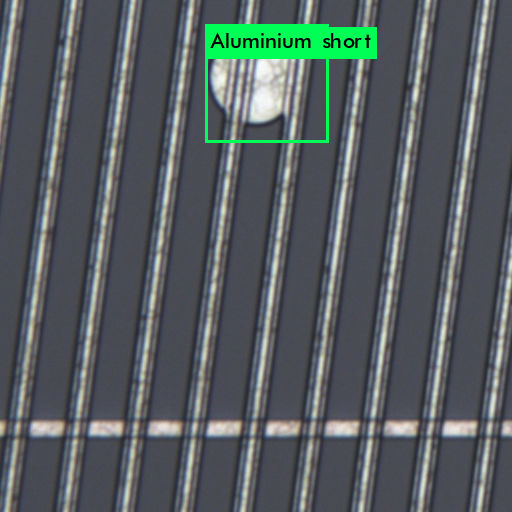}
\caption{Various surface defects on the silicon sensor identified using convolutional deep neural network.}
%\end{center}
\label{fig:defects}
\end{figure*}

In our work, we present a full statistical analysis of the various defect types observed on the surface of 1191 sensors scanned in an automated fashion on a dedicated inspection setup and analyzed with advanced methods using Computer Vision (CV) and Convolutional Deep Neural Networks (CDNNs) \cite{yolo}. We also report on the sensors' metrological measurements, i.e., warp measurement of the various sensor variants. We believe that this comprehensive analysis can help optimizing future QA procedures and inspection protocols for similar applications.

The surface scanning of every sensor was performed using a custom made, automated optical inspection setup shown in Figure \ref{fig:Setup}. The sensor is placed on the XY-stage which is connected to the vacuum system to hold the sensors during their inspection. The XY-stage can move linearly with the help of Fauhaber/Movtec$^{\circledR}$ SMC-300 servo motors within a range of 200 mm and 70 mm in X and Y direction
%\vspace*{3.9cm} 
respectively, with a precision 0.3 $\mu$m. 
Detailed technical information about the setup, sensor alignment procedure and the scanning process has already been discussed in \cite{Lavrik, lavrik2}.

Figure \ref{fig:surface_defects} shows the surface defects on the silicon sensors which were used as ground truth for the training of the neural network. The neural network used in the training the model is based on the Darknet implemented with the modified R-CNN framework model. It also allows to classify the object and to locate them in the images against the trained data. The key features of the neural network are convolutional layers, kernels, activation functions etc \cite{Lavrik, FRNN, kaiming, Nwankpa}.  
Figure \ref{fig:defects} depicts various surface defects detected by the trained neural network. The efficiency of the CDNN to locate the defects is achieved to be 91.5 $\%$ with an error of nearly 8.5 $\%$. 
The equation \ref{eq:accuracy} and \ref{eq:error}  are used to calculate the efficiency and error of the CDNN respectively.

\begin{center}
%\begin{strip}
\begin{equation} \label{eq:accuracy}
    Accuracy =  1 - Error
\end{equation}
%\end{strip}

\begin{strip}
\begin{equation} \label{eq:error}
   Error = \frac{False ~Positive + False ~Negative} {True ~Positive + True ~Negative + False ~Positive + False ~Negative} 
\end{equation}
\end{strip}
\end{center}

\section{Results}\label{se:results}

In this section, we present the statistics of defective strips, and the analysis of the quality grade and quality score based on the total amount of defect detected on the sensor's surface. Also, defective strip maps and defective strip cluster analysis and mapping of the defects of a particular class based on their positions on the sensor's surface are presented.

\subsection{Total defective strips}

As per the agreement with Hamamatsu Photonics, up to 30 defective strips (aluminium defects and pinholes only) per sensor were allowed for the devices to be accepted by their quality control.
However, it is conceivable that there are 
more defective strips on the sensor's surface which had not been detected by the vendor's QA. 

Therefore, to ensure the sensor's integrity and guarantee its prescribed performance it is required to pay attention to the total amount of defective strips of all kinds on the sensor's surface. For counting the total number of bad strips, only four defects, such as implant break, p-stop break, aluminium open and aluminium short were considered because they lead to a faulty read-out channel. Also their presence in large amounts on consecutive strips could create a dead zone on the sensor and the particles traversing the sensor in such a dead zone could potentially not be detected.

\begin{figure*}
\centering
\includegraphics[height = 6.7cm, width = 7.4cm]{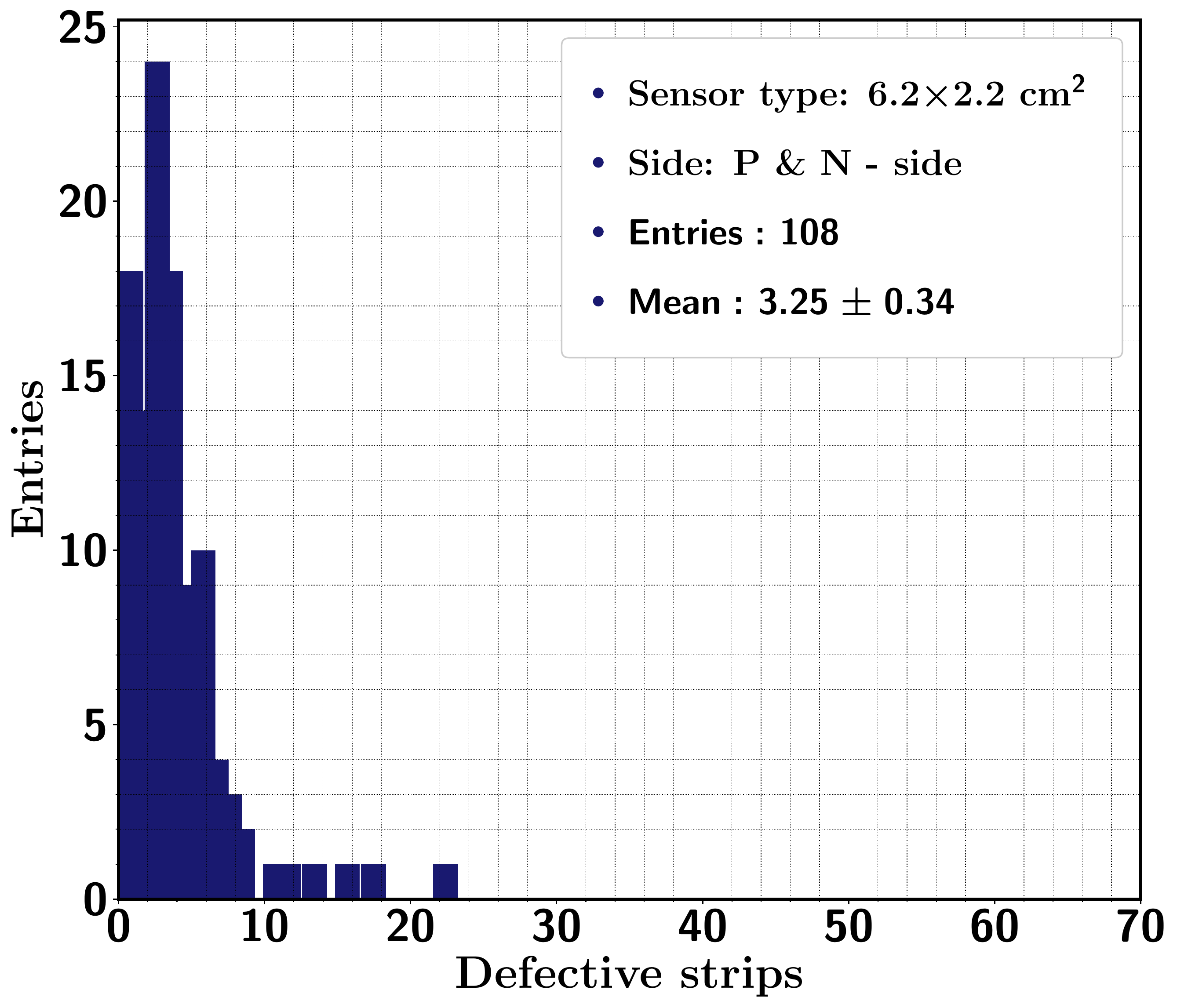}
\includegraphics[height = 6.7cm, width = 7.4cm]{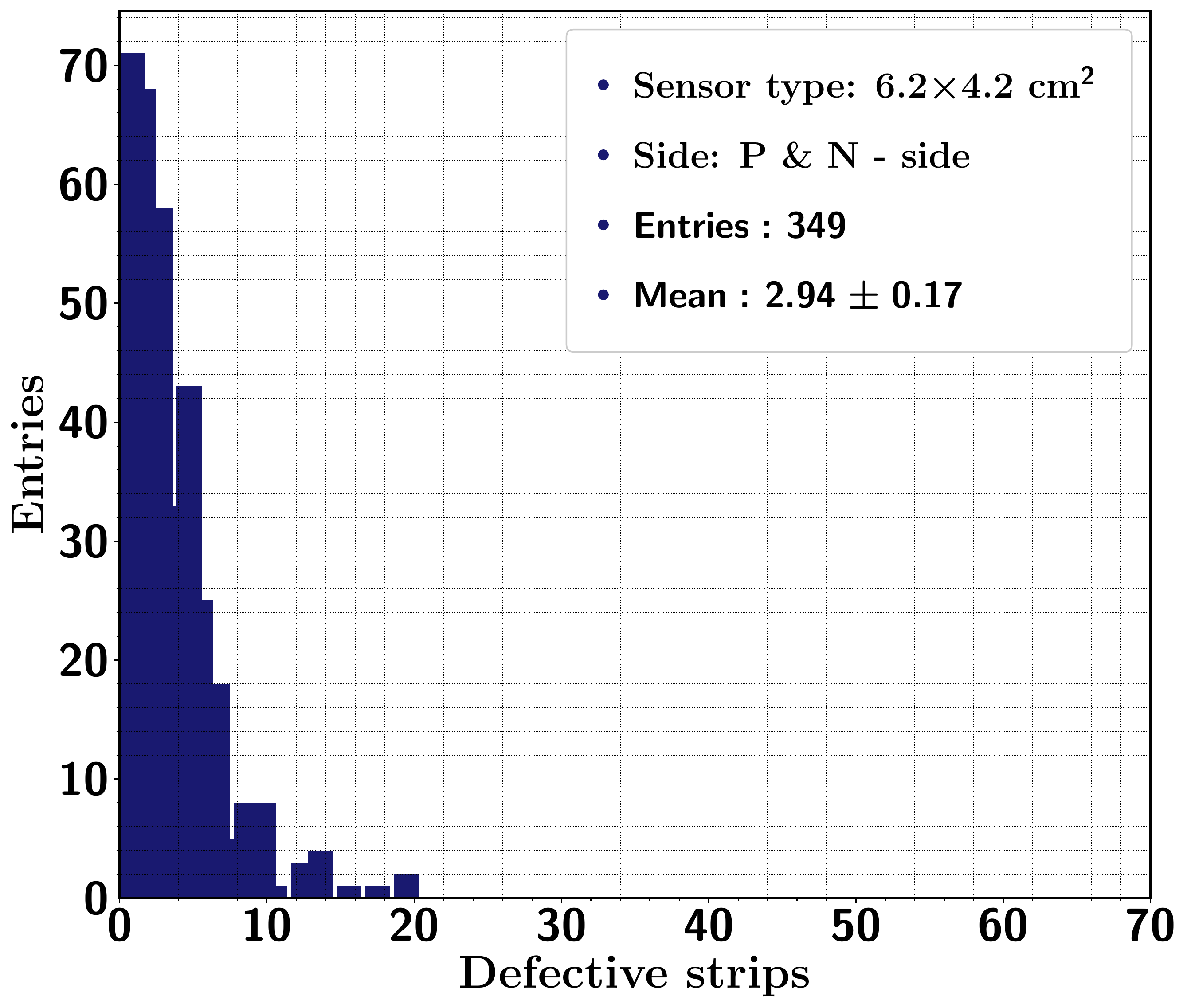}
\includegraphics[height = 6.7cm, width = 7.4cm]{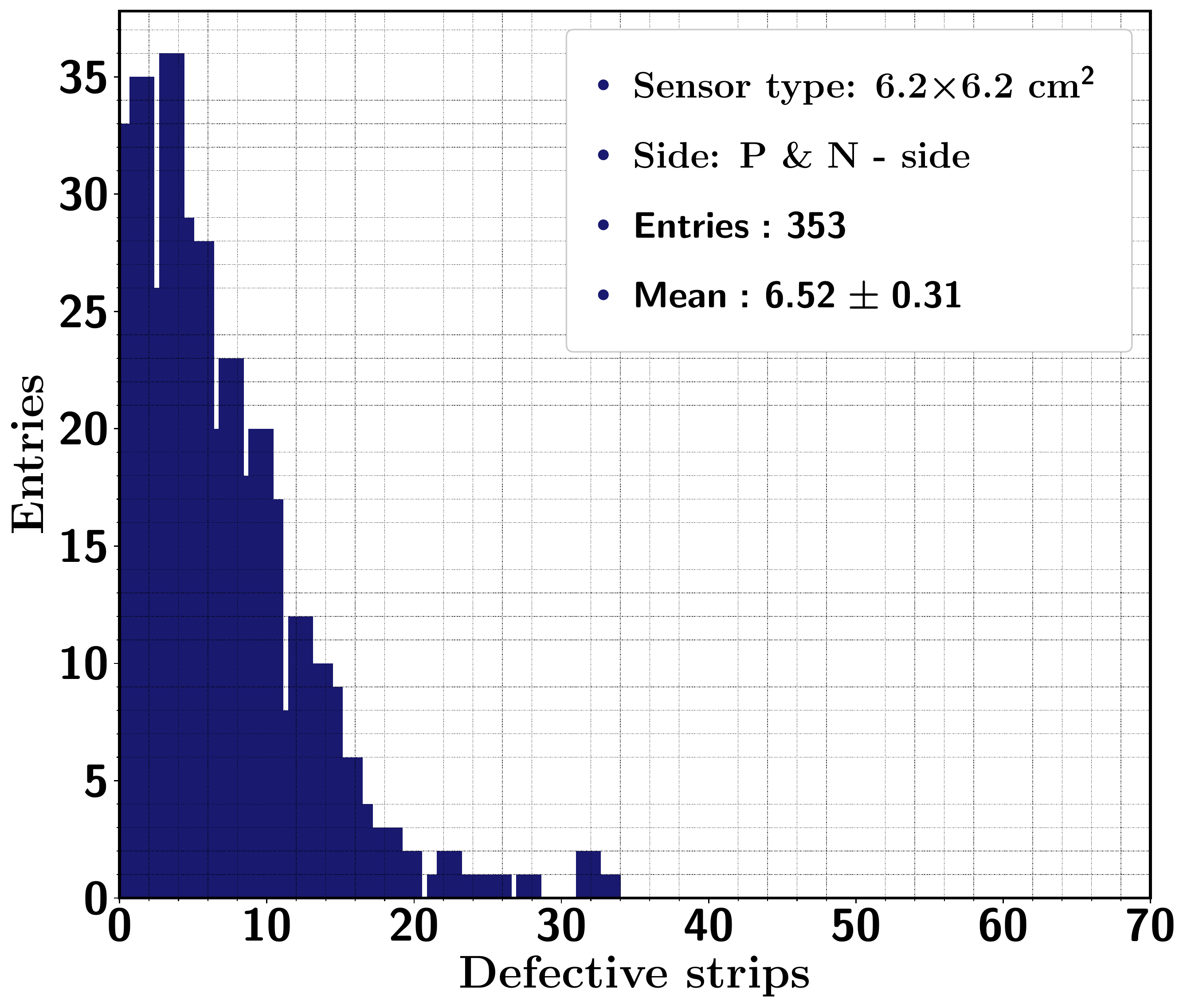}
\includegraphics[height = 6.7cm, width = 7.4cm]{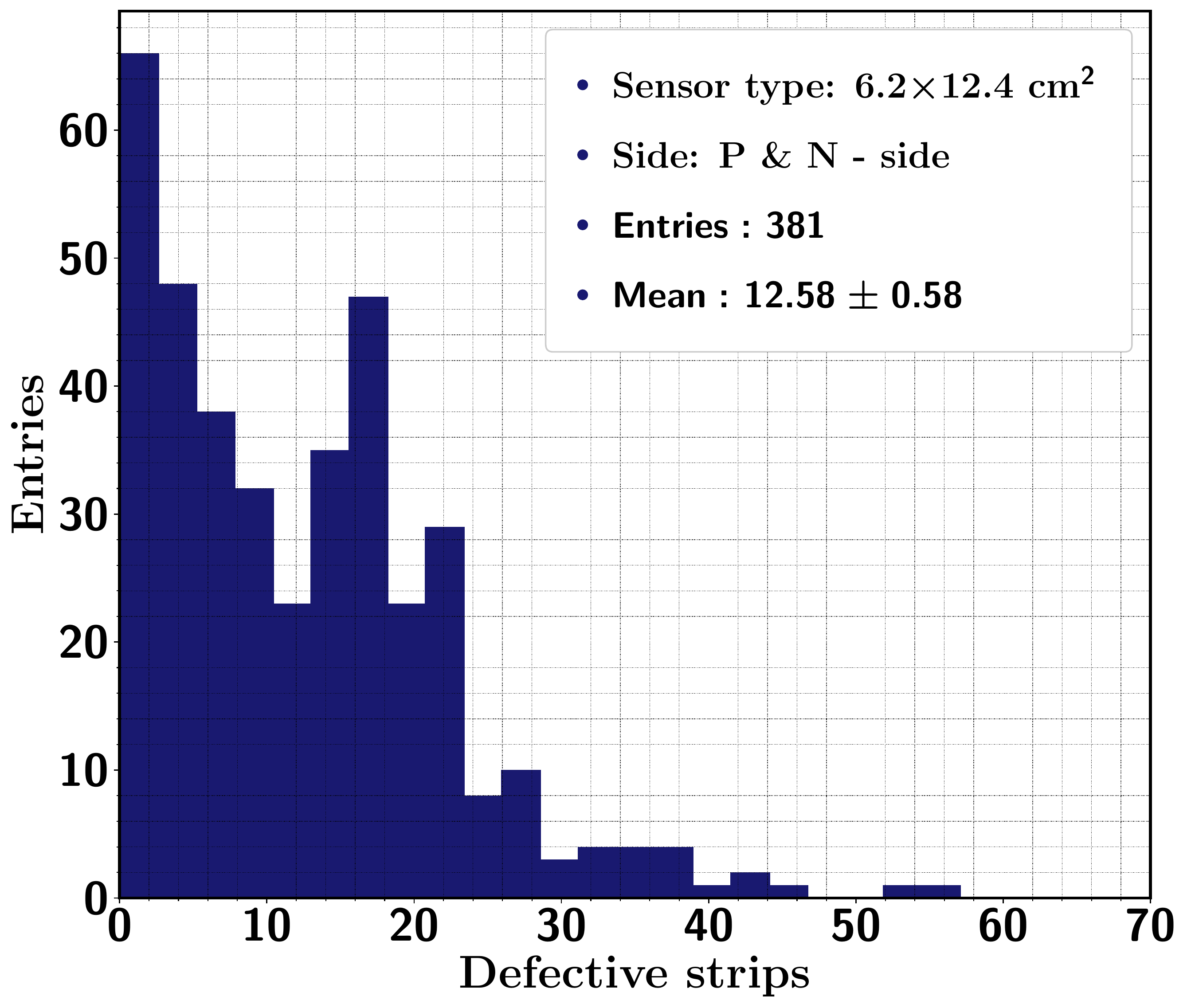}
\caption{Distribution of the total number of defective strips per sensor based on the detected defects.}
\label{fig:Total_defective_strips1}
\end{figure*}

Figure \ref{fig:Total_defective_strips1} shows the total number of defective strips for individual sensor variants. It is observed that for the size 6.2~$\times$ 2.2 cm$^{2}$ and 6.2 $\times$ 4.2 cm$^{2}$, most of the sensors have less than 15 defective strips per sensor. On the other hand, for 6.2 $\times$ 6.2 cm$^{2}$ and 6.2 $\times$ 12.4 cm$^{2}$ sizes many sensors were found with more than 15 defective strips.

\subsection{Quality grade and Quality score}
For the sensor selection and proper geometrical arrangement, which in turn ensures the best particle tracking in the final tracking system, we introduce two metrics. The first one, the quality grade, is based on the overall amount of the defective strips found on a sensor. The second one, the quality score, takes into account all kinds of defects found on the sensor and is used to further rank the sensors within the same quality grade.

For the quality grading scheme, sensors are classified into four different grade classes by counting the total number of defective strips of the following types: implant break, p-stop break, aluminium open and aluminium shorts. The grading scheme is classified into four categories A, B, C and D. The number of defects allowed per sensor in each grade is presented in Table \ref{tab:defect}.  The sensors in a grade A category are those with very less number of defects, while those sensors in grade D have large amount of defects. The resulting grade breakdown for each sensor size is shown in Table \ref{tab:sensor_grade}.

\begin{table}[ht!]
\begin{adjustbox}{width=7cm, center}
%\resizebox{\columnwidth}{!}{%
\begin{tabular}{l*{4}{c}r}\hline
& \multicolumn{4}{c}{Quality grade} \\
\hline
    & { A } & { B } & { C }&
    {D}\\
\hline \hline 
Defects  & $\leq$ 5 & $\leq$ 15 & $\leq$ 25 & $ > $ 25 &
\\
\hline 
\end{tabular}  
\end{adjustbox}
\caption{Amount of defect within each of the four grades.}
\label{tab:defect}
\end{table}

\begin{table}[ht!]
\begin{adjustbox}{width=7.3cm, center}
%\resizebox{\columnwidth}{!}{%
\begin{tabular}{l*{4}{c}r}\hline
 & \multicolumn{4}{c}{Quality grade} \\
\hline
Sensor size (cm$^{2}$) & A  & B & C & D \\
\hline \hline \\[-1mm]
6.2 $\times$ 2.2 & 93 & 13 & 2 & 0\\[2mm]
6.2 $\times$ 4.2 & 296 & 50 & 3 & 0\\[2mm]
6.2 $\times$ 6.2 & 185 & 144 & 19 & 5 &\\[2mm]
6.2 $\times$ 12.4 & 114 & 125 & 107 & 35& \\
\hline \\[-1.7mm]
All & 688 & 332 & 131 & 40 &\\
\hline\\
\end{tabular}
\end{adjustbox}
\caption{Breakdown of quality grade by sensor size. }
\label{tab:sensor_grade}
\end{table}

It has been found that most of the sensors, 688 specimens, fall in grade A. This means those sensors in the quality grade A have a minimum amount of the defects and can be safely used for module production.
Moreover, 332 and 131 sensors were found to be in the sensor grades B and C, respectively. Only 40 sensors of the size 6.2 $\times$ 6.2 cm$^{2}$ and 6.2 $\times$ 12.4 cm$^{2}$ were observed in grade D.

%\subsection{Quality score}

The quality score for every sensor was determined, normalized to the value between 0 and 1, by counting all the possible defect types present on the sensor's surface and applying weighted penalty factors to each of those defect types.
Quality score of 0 means the sensors have high number of surface defects, whereas sensors with a quality score of 1 have a lower amount of surface defects.

\begin{figure*}
\centering
\includegraphics[height = 6.7cm, width = 7.4cm]{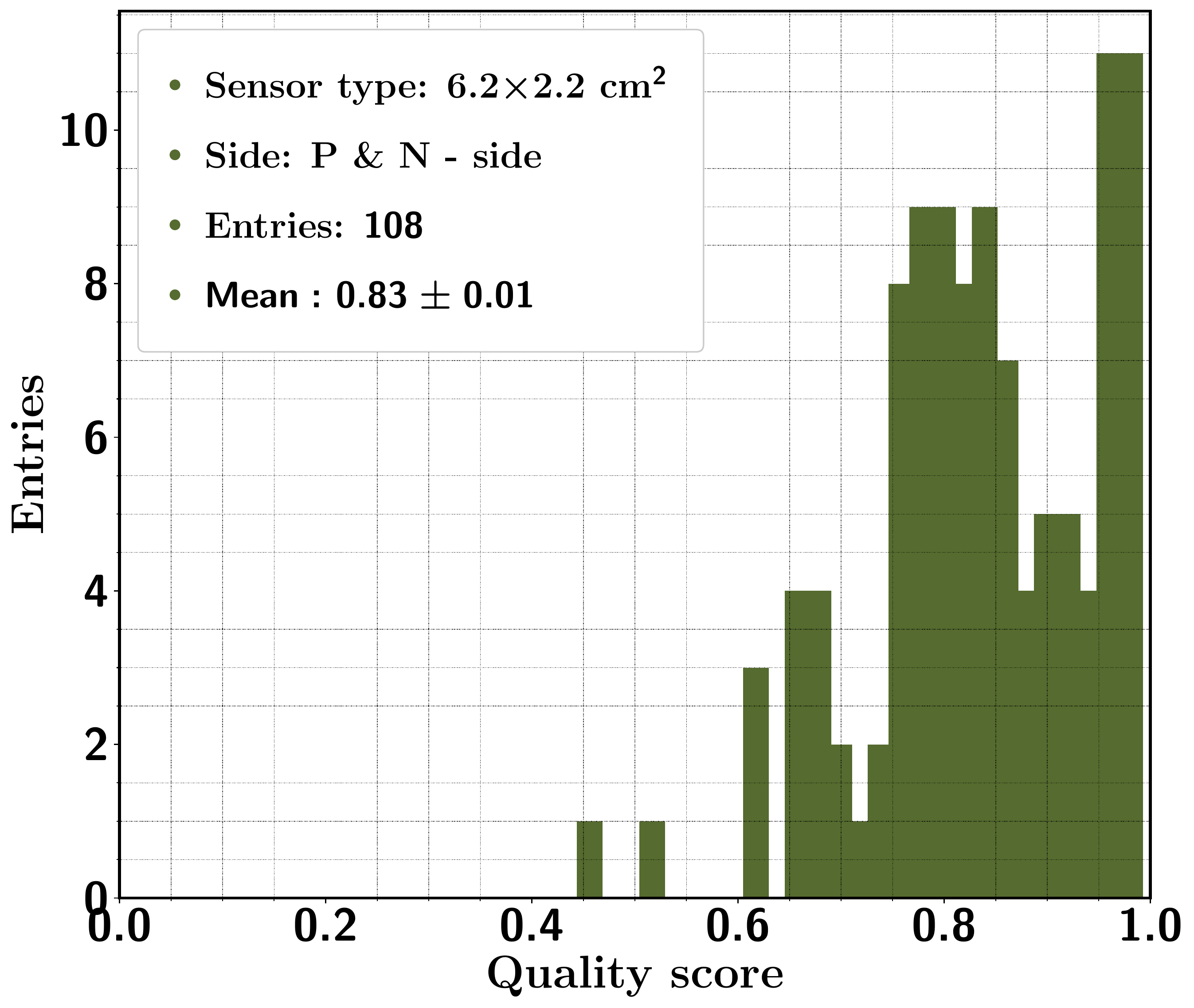}
\includegraphics[height = 6.7cm, width = 7.4cm]{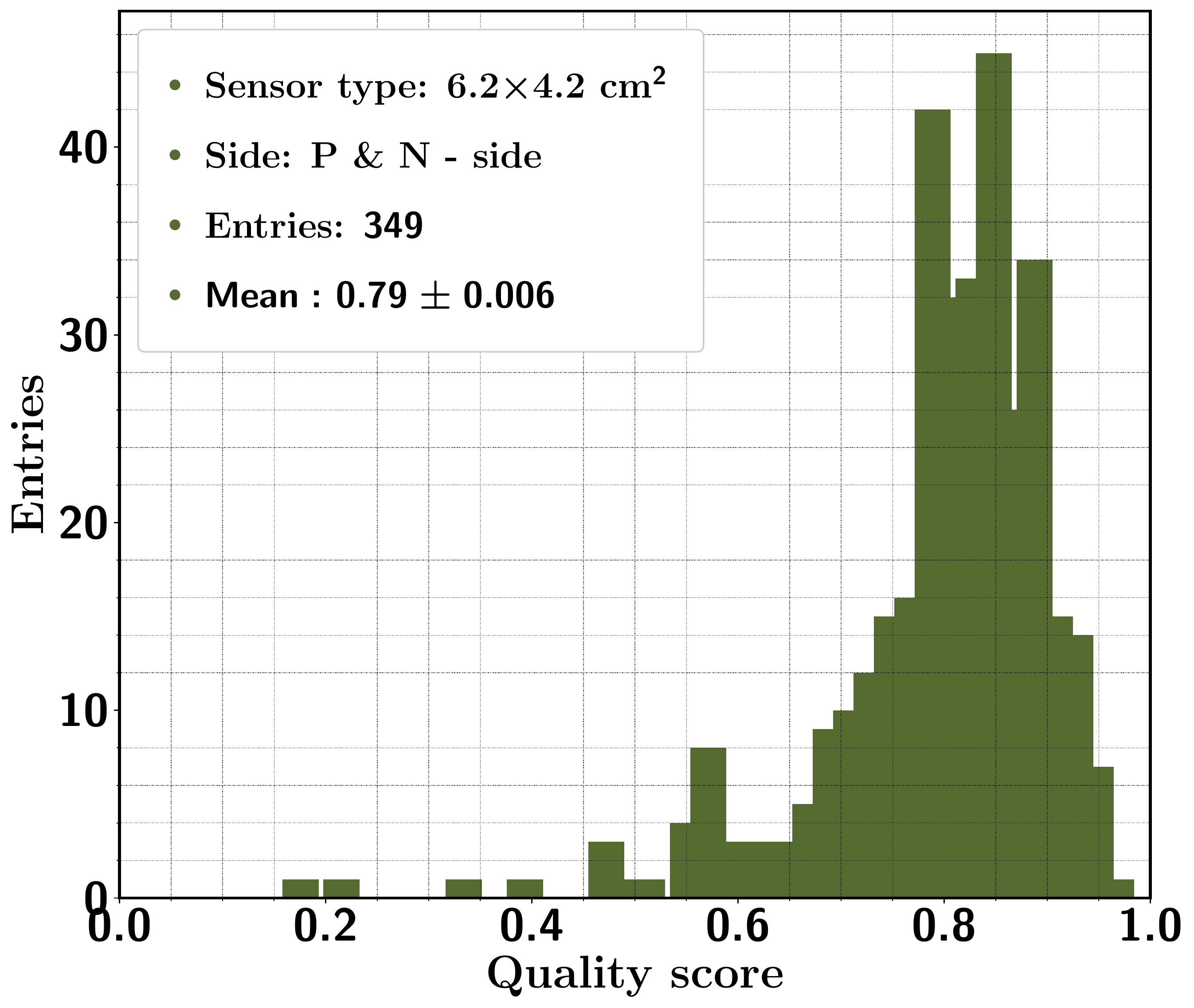}
\includegraphics[height = 6.7cm, width = 7.4cm]{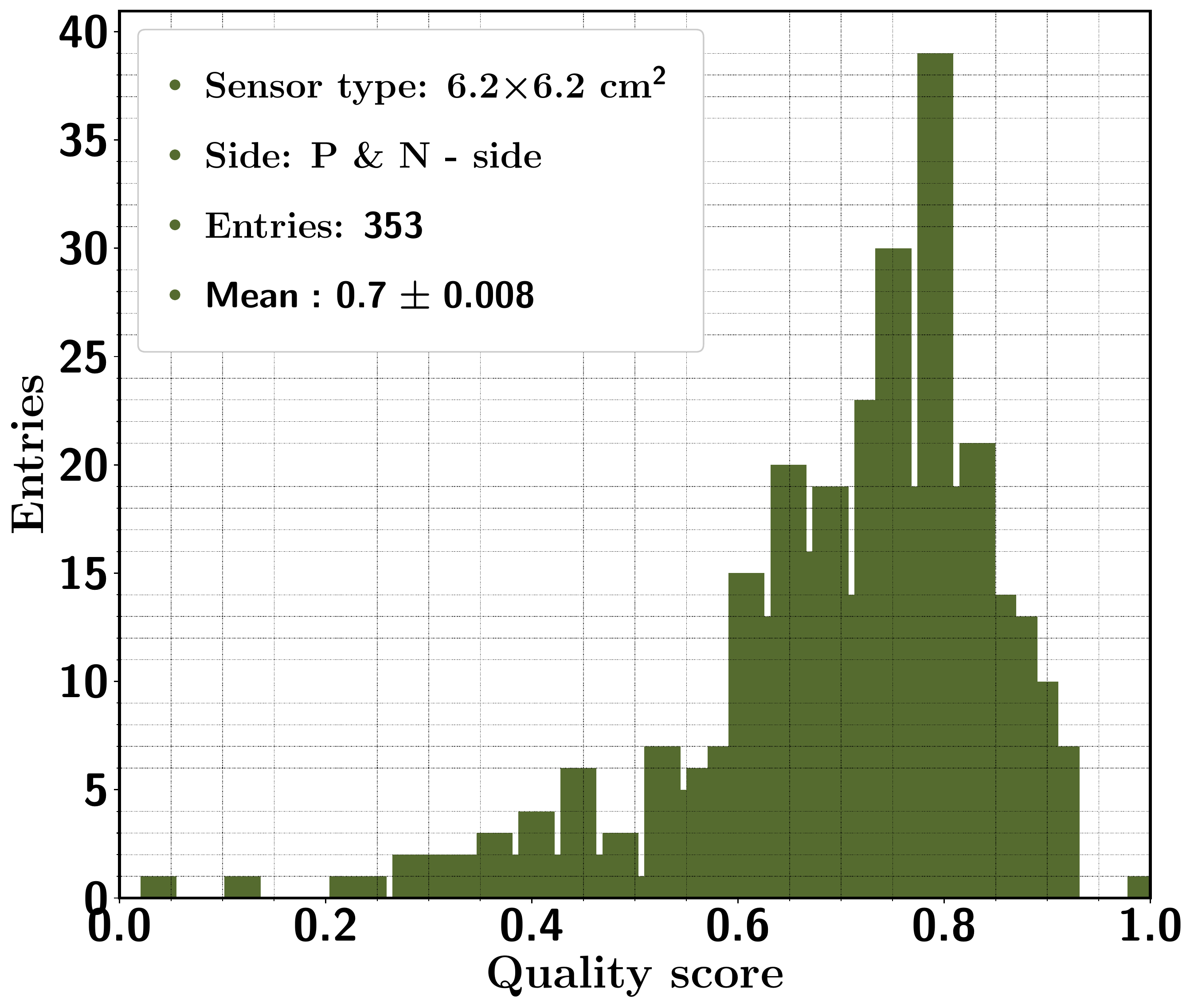}
\includegraphics[height = 6.7cm, width = 7.4cm]{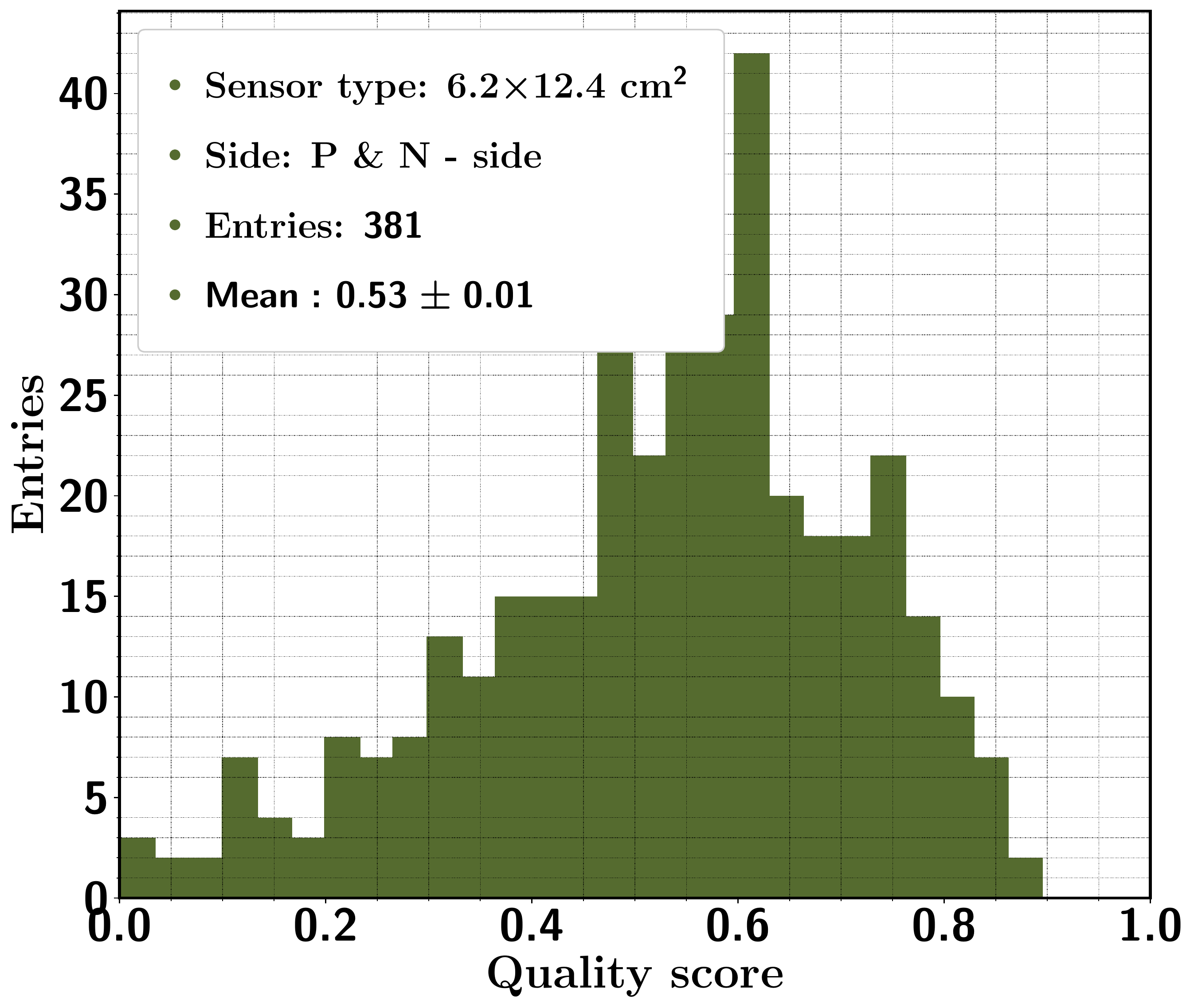}
\caption{Quality score of the sensor variants 6.2 $\times$ 2.2 cm$^{2}$,  6.2 $\times$ 4.2 cm$^{2}$ (Top panel) and 6.2 $\times$ 6.2 cm$^{2}$, 6.2 $\times$ 12.4 cm$^{2}$ (Bottom panel).}
\label{fig:Quality_score}
\end{figure*}

As can be seen from the Figure \ref{fig:Quality_score}, 
the mean value of the quality score for each sensor variant was attained as 83$\%$, 79$\%$, 70$\%$ and 53$\%$ respectively. %From this, we conclude that
Apparently, the quality score is decreasing with increasing sensor size, with 
the amount of defects being evidently proportional to the sensor's surface area.

\subsection{Defective strip maps}
\label{sec:defective_strip_map}
There are 1024 strips on each side of the silicon sensor. The defective strip map represents the distribution of defected strips on the sensor's surface.
\begin{figure*}
\centering
\includegraphics[height = 7.9cm, width = 7.15cm]{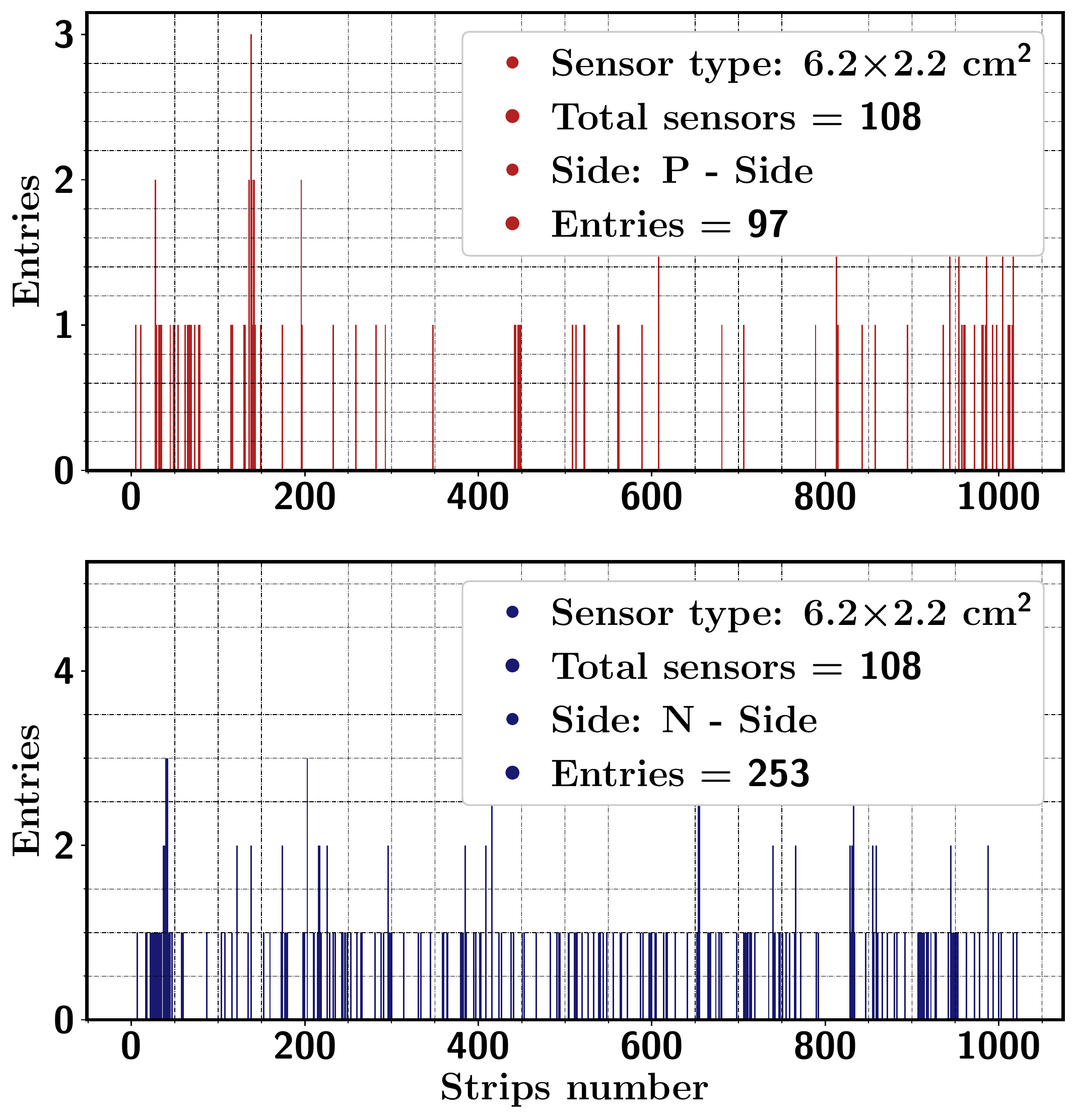}
\includegraphics[height = 7.9cm, width = 7.15cm]{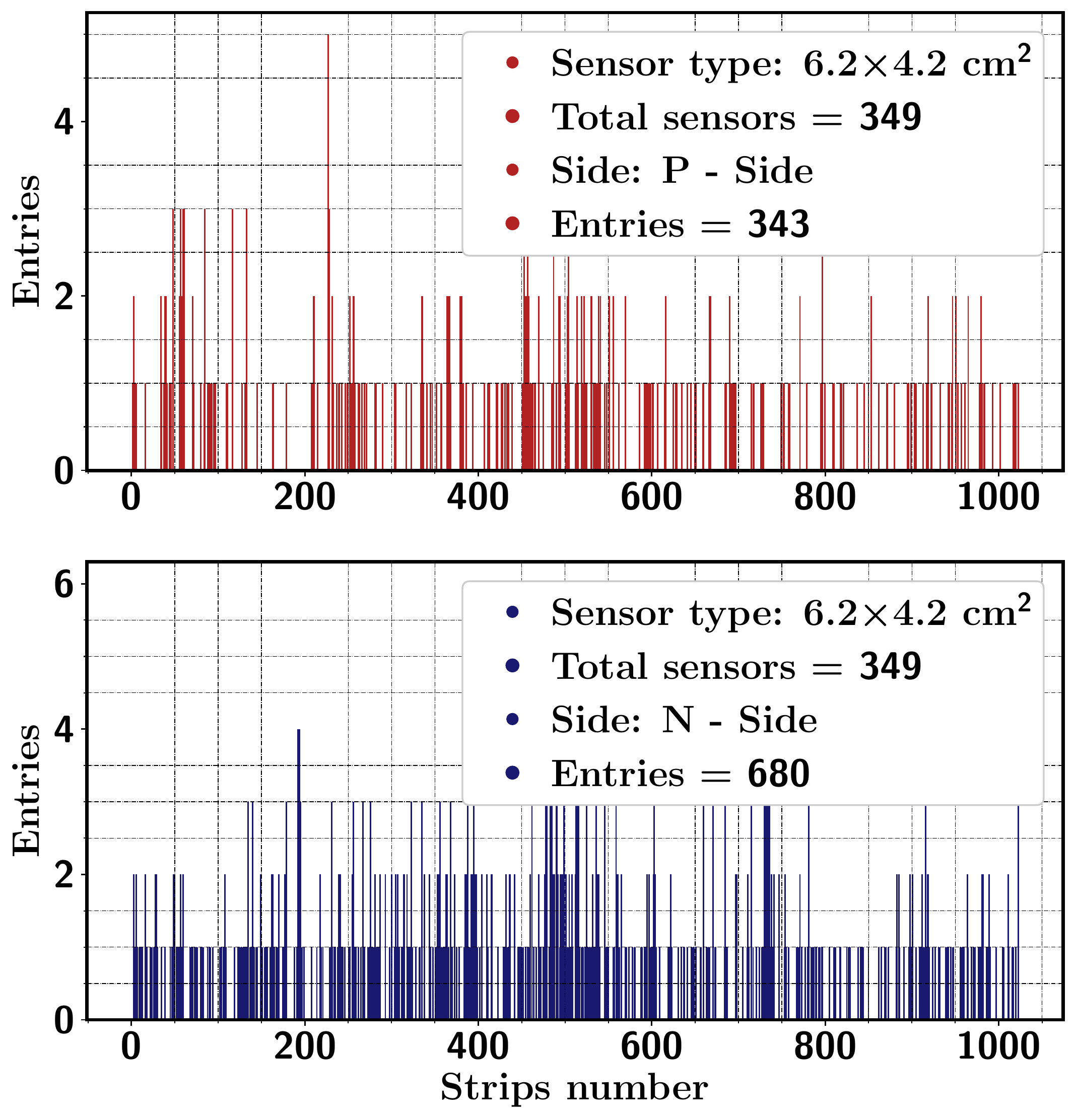} \\[5mm]
\includegraphics[height = 7.9cm, width = 7.15cm]{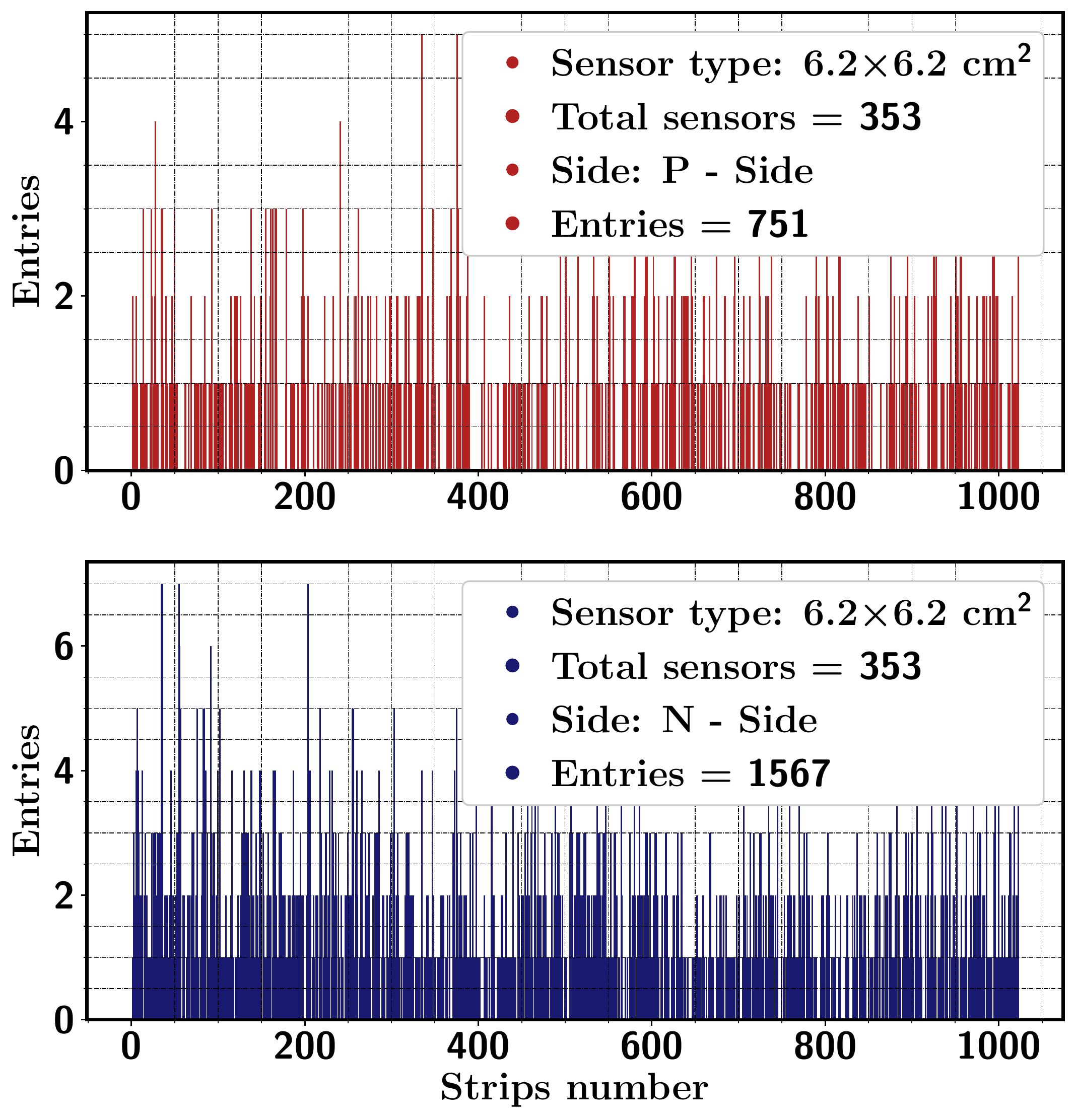}
\includegraphics[height = 7.9cm, width = 7.15cm]{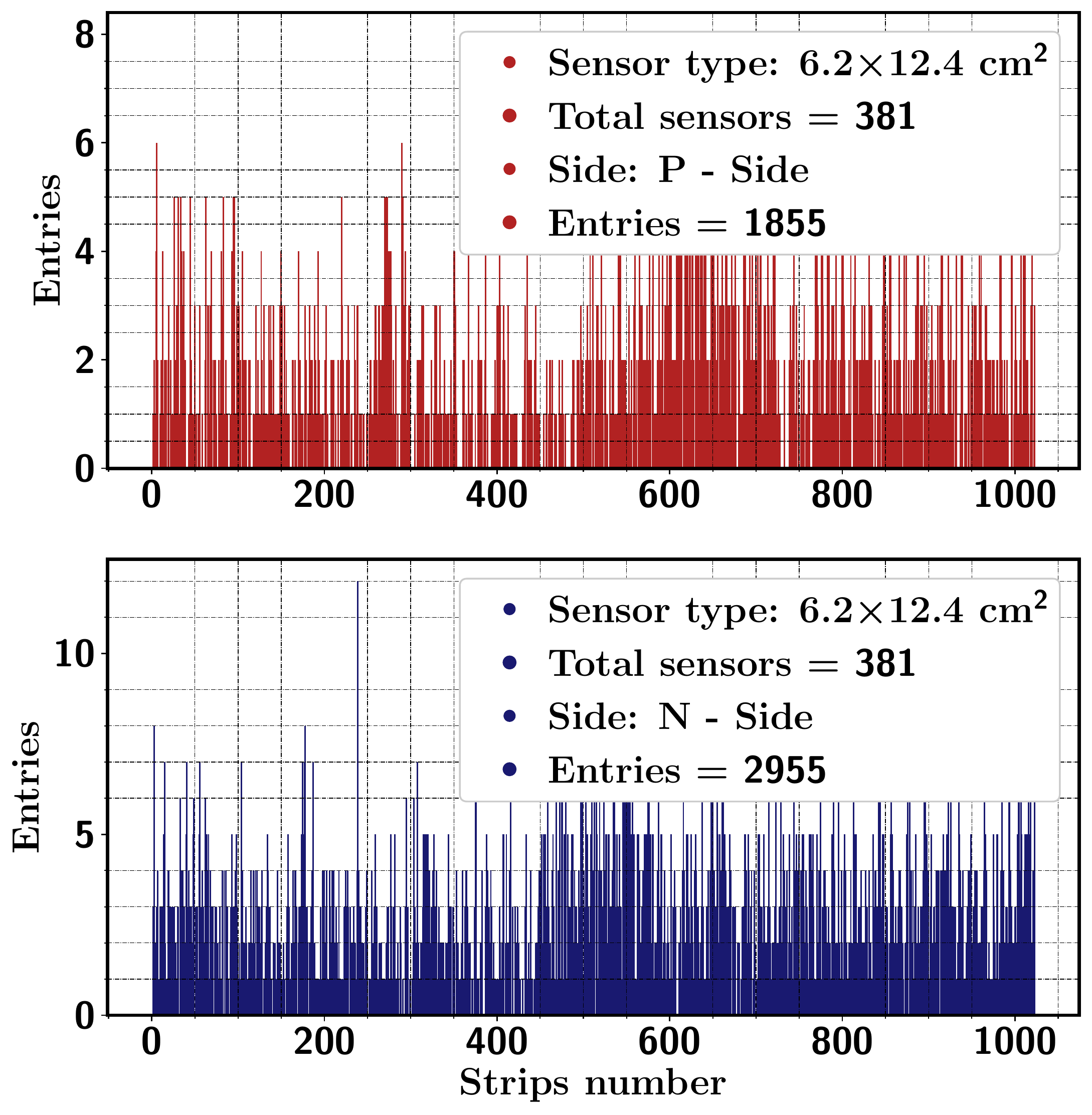}
\caption{Accumulated defective strip maps for both p-side and n-side of the sensor variants 6.2 $\times$ 2.2 cm$^{2}$, 6.2 $\times$ 4.2 cm$^{2}$ and 6.2 $\times$ 6.2 cm$^{2}$, 6.2 $\times$ 12.4 cm$^{2}$.}
\label{fig:Defect_strip_maps1}
\end{figure*}
Similarly to the quality grade, only the defective strips of implant break, p-stop break, aluminium open and aluminium shorts were taken into account.

Figure \ref{fig:Defect_strip_maps1} shows the defective strip maps, individually for both p-side and  n-side, of the sensor variants 6.2 $\times$ 2.2 cm$^{2}$, 6.2 $\times$ 4.2 cm$^{2}$, 6.2 $\times$ 6.2 cm$^{2}$ and 6.2 $\times$ 12.4 cm$^{2}$.
From the defective strip maps, it is evident that the number of defective strips on the sensor's p-side was found to be less than that of the n-side for each sensor variants. On the other hand, one can deduce that the distribution of defective strips is relatively uniform without any particular region on the sensor's surface with a high number of defective strips. However, there was a substantial amount of clusters of defective strips found.
 %on the sensor's surface.

\subsection{Defective strip clusters}
\label{sec:defective_strip_cluster}

\begin{figure*}
\centering
\includegraphics[height = 6.2cm, width = 7.2cm]{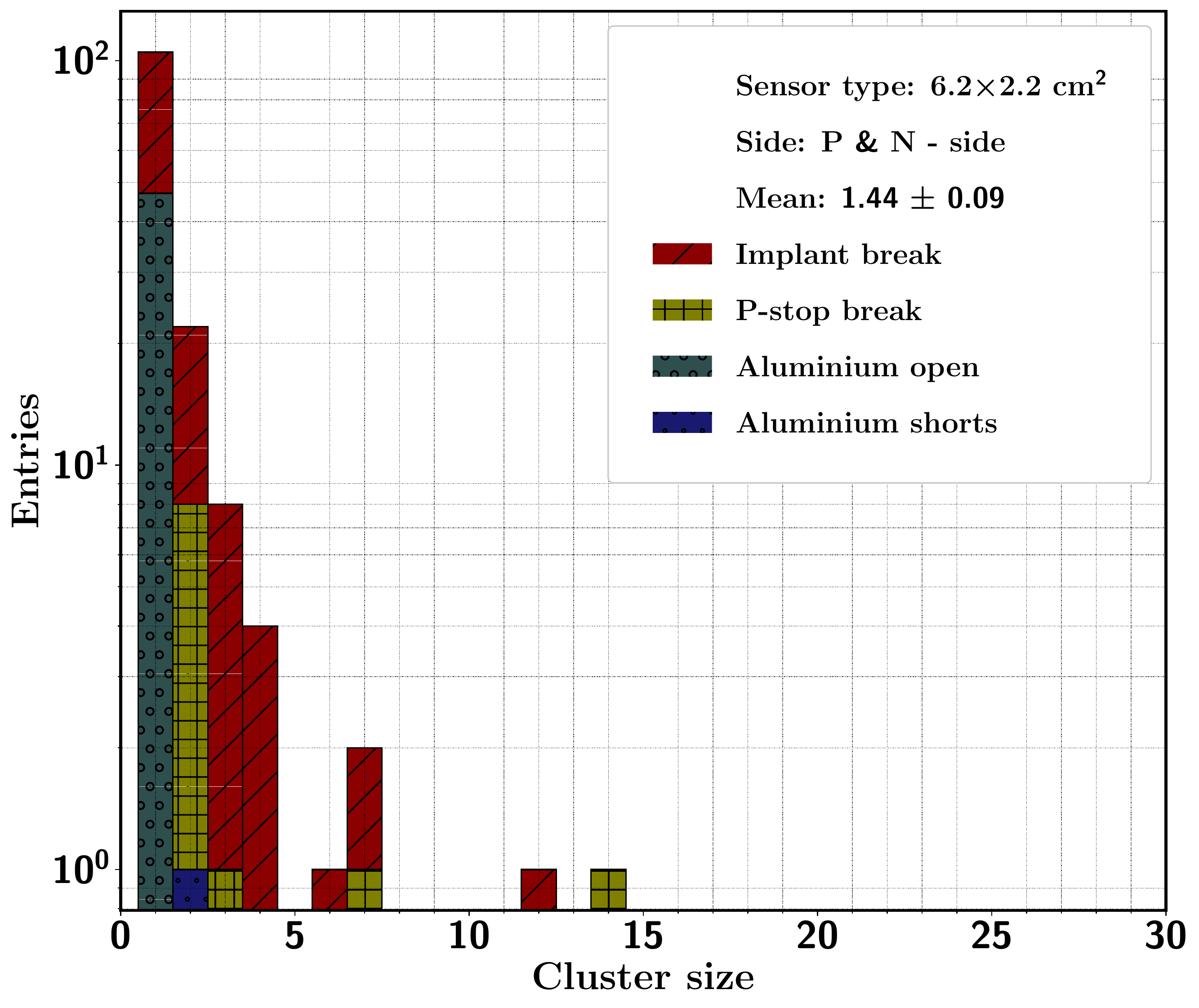}
\includegraphics[height = 6.3cm, width = 7.2cm]{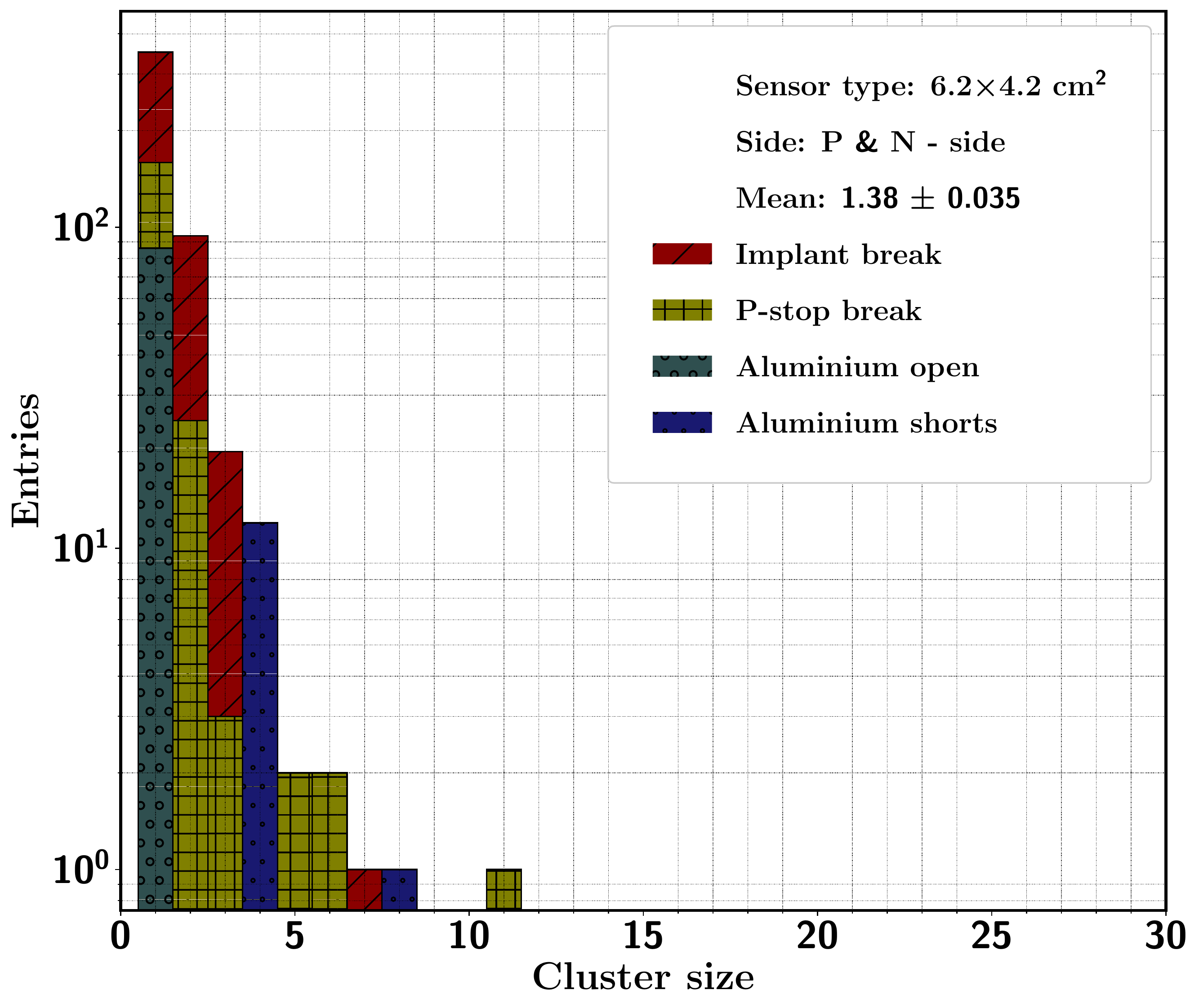}
\includegraphics[height = 6.2cm, width= 7.2cm]{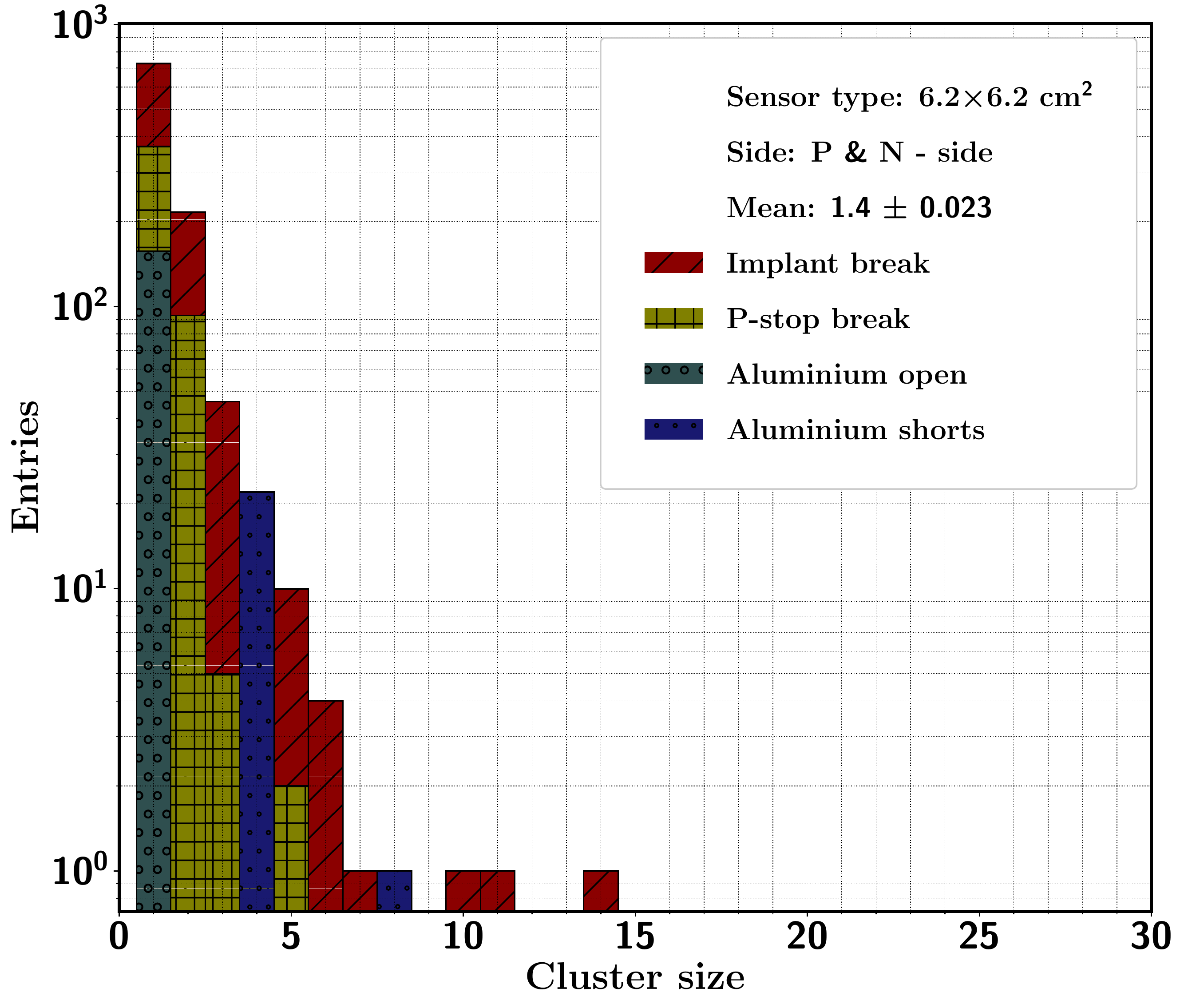}
\includegraphics[height = 6.2cm, width = 7.2cm]{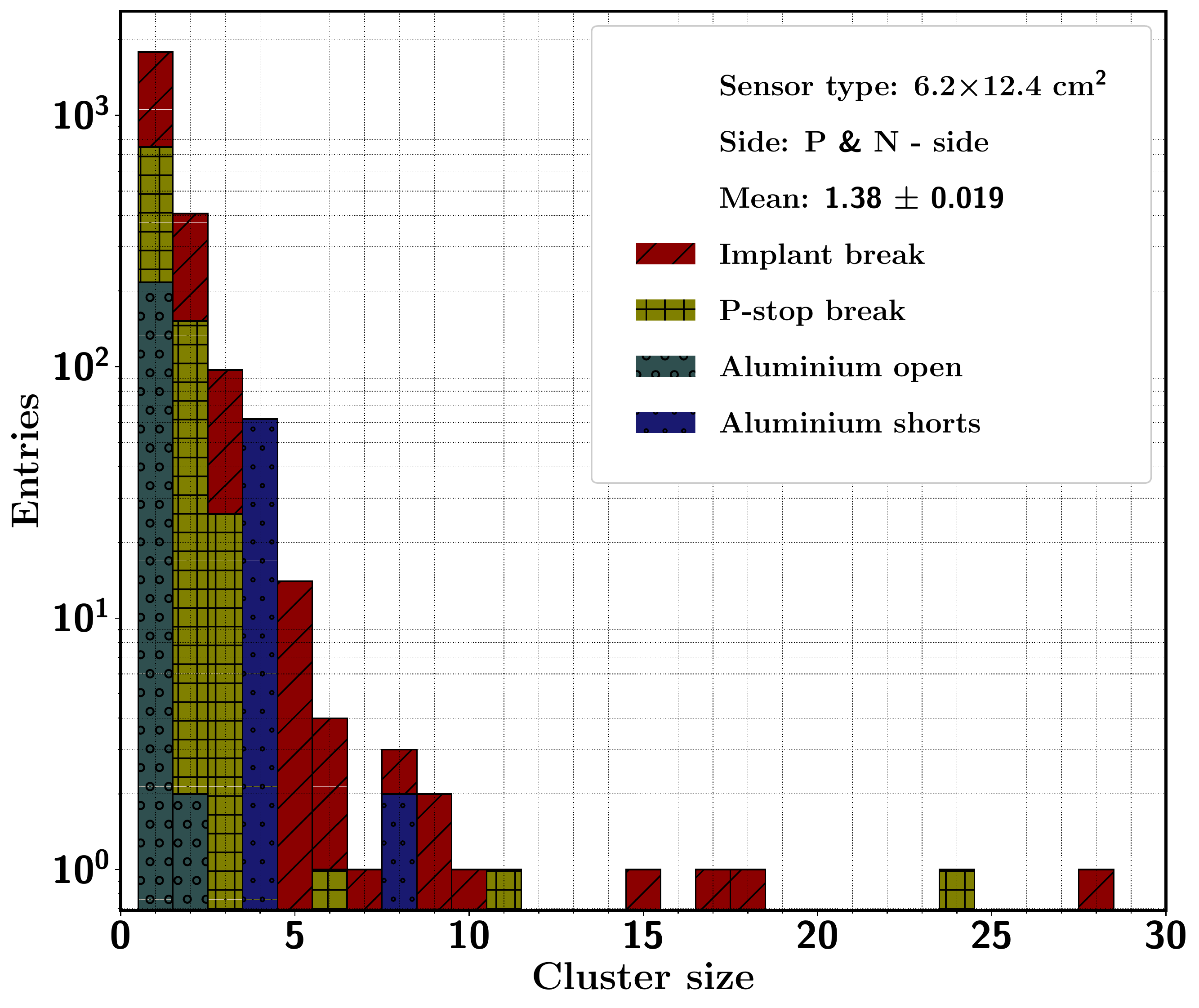}
\caption{Defective strip clusters observed on p- and n-sides of sensor variants 6.2 $\times$ 2.2 cm$^{2}$, 6.2 $\times$ 4.2 cm$^{2}$ (Top panel) and 6.2 $\times$ 6.2 cm$^{2}$, 6.2 $\times$ 12.4 cm$^{2}$(Bottom panel). }
\label{fig:Defect_cluster}
\end{figure*}

From defective strip maps shown in Section \ref{sec:defective_strip_map}, it is clear that some clusters of defective strips are visible on the sensor's surface. Therefore, it is interesting to know how these clusters are sized and distributed.

For finding defective strip clusters, only the defects of implant break, p-stop break, aluminium open, aluminium shorts on the consecutive strips were taken into consideration in the algorithm.
Figure \ref{fig:Defect_cluster} illustrates the defective strip clusters for the sensor variants 6.2 $\times$ 2.2 cm$^{2}$, 6.2 $\times$ 4.2 cm$^{2}$, 6.2 $\times$ 6.2 cm$^{2}$ and 6.2 $\times$ 12.4 cm$^{2}$ respectively.  It was found that the clusters of aluminium open, aluminium shorts defects are of negligible amount on the surface of all four sensor variants. Furthermore, only on a few sensors of the sensor size 6.2 $\times$ 2.2 cm$^{2}$, 6.2 $\times$ 4.2 cm$^{2}$ and 6.2 $\times$ 6.2 cm$^{2}$ we have found clusters of more than 5 implant breaks and p-stop breaks defective strips.  Many 6.2 $\times$ 12.4 cm$^{2}$ sensors were found with more than five consecutive implant breaks and p-stop-breaks defective strips. Among all, the biggest cluster of implant breaks (28 strips) and p-stop breaks (24 strips) was found on a 6.2 $\times$ 12.4  cm$^{2}$ sensor.

\subsection{2-D map of defects}

\begin{figure*}
\centering
\includegraphics[height = 9.25cm, width = 7.4cm]{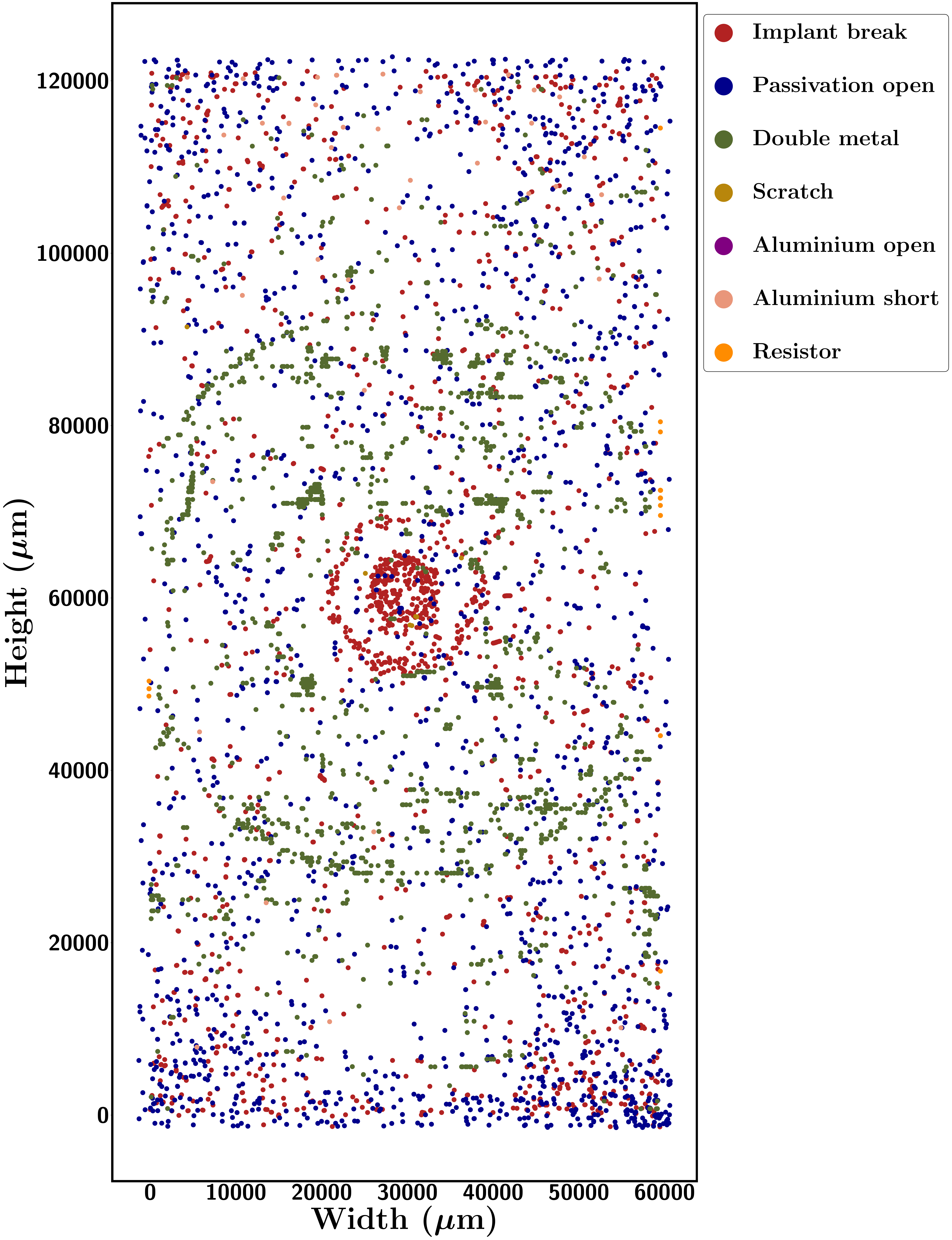}
\includegraphics[height = 9.25cm, width = 7.4cm]{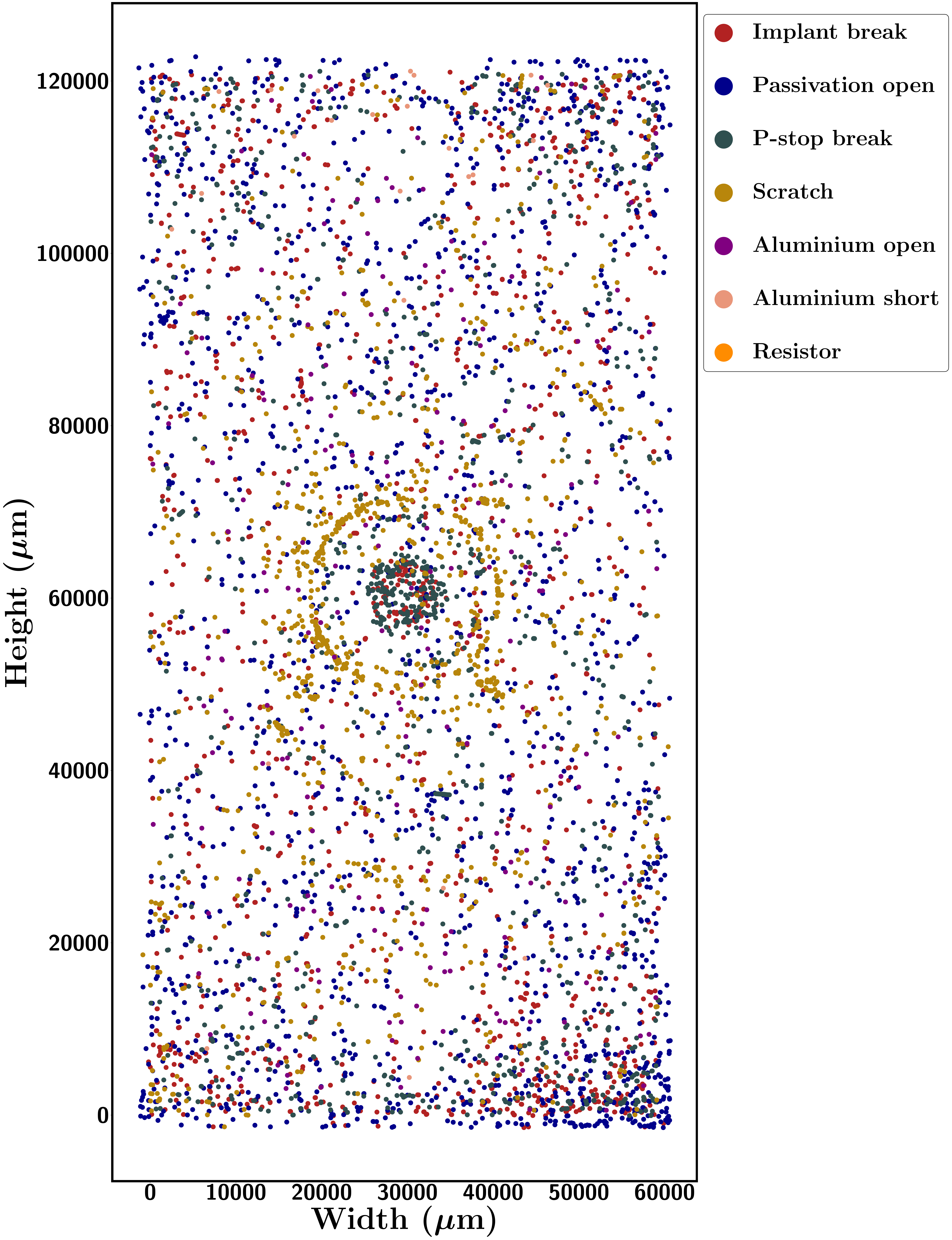}
\caption{Defects map of the sensor variant 6.2 $\times$ 12.4 cm$^{2}$ for p-side (Left)  and n-side (Right). }
\label{fig:Defect_map_6_12_P_side}
\end{figure*}

A two-dimensional map of the defects provides us with a clear view of their actual position on the sensor surface. The 2D map of defects was obtained by converting the local pixel coordinate 
to its global coordinate.

Figure \ref{fig:Defect_map_6_12_P_side} shows the 2D defect map individually for p- and n-side of the sensor variant 6.2 $\times$ 12.4 cm$^{2}$.
It is striking that some defects like double metal line and implant break are found to be forming a circular pattern. In a similar shape, the defects like scratch, p-stop and implant break defects are observed on the n-side of the sensor's surface.
The pattern is attributed to the usage of a vacuum tool to handle the sensor during production at the production site.

\section{Warp measurement}

Metrological measurements such as warp measurement provide information about the actual geometry and irregularities of the sensor's surface. In this section, we present the warp measurement results obtained for different sensor variants using the same custom made optical inspection setup as shown in Figure \ref{fig:Setup}.  It uses the focus variation technique for the height profile measurement of the object under test. The conversion ratio motor steps to micrometer for the focus motor stage was used to convert the values of height profile obtained in motor steps to real-world measurement. Further details about the method for contactless metrology measurement has been presented in \cite{lavrik2,Warp}.
The precise knowledge on the sensors warp is an important information during the module assembly procedure, where we aim for an alignment precision of the sensors within $\pm$~100~µm.

\begin{figure}[ht!]
\centering
\includegraphics[height=6.1cm, width = 8cm]{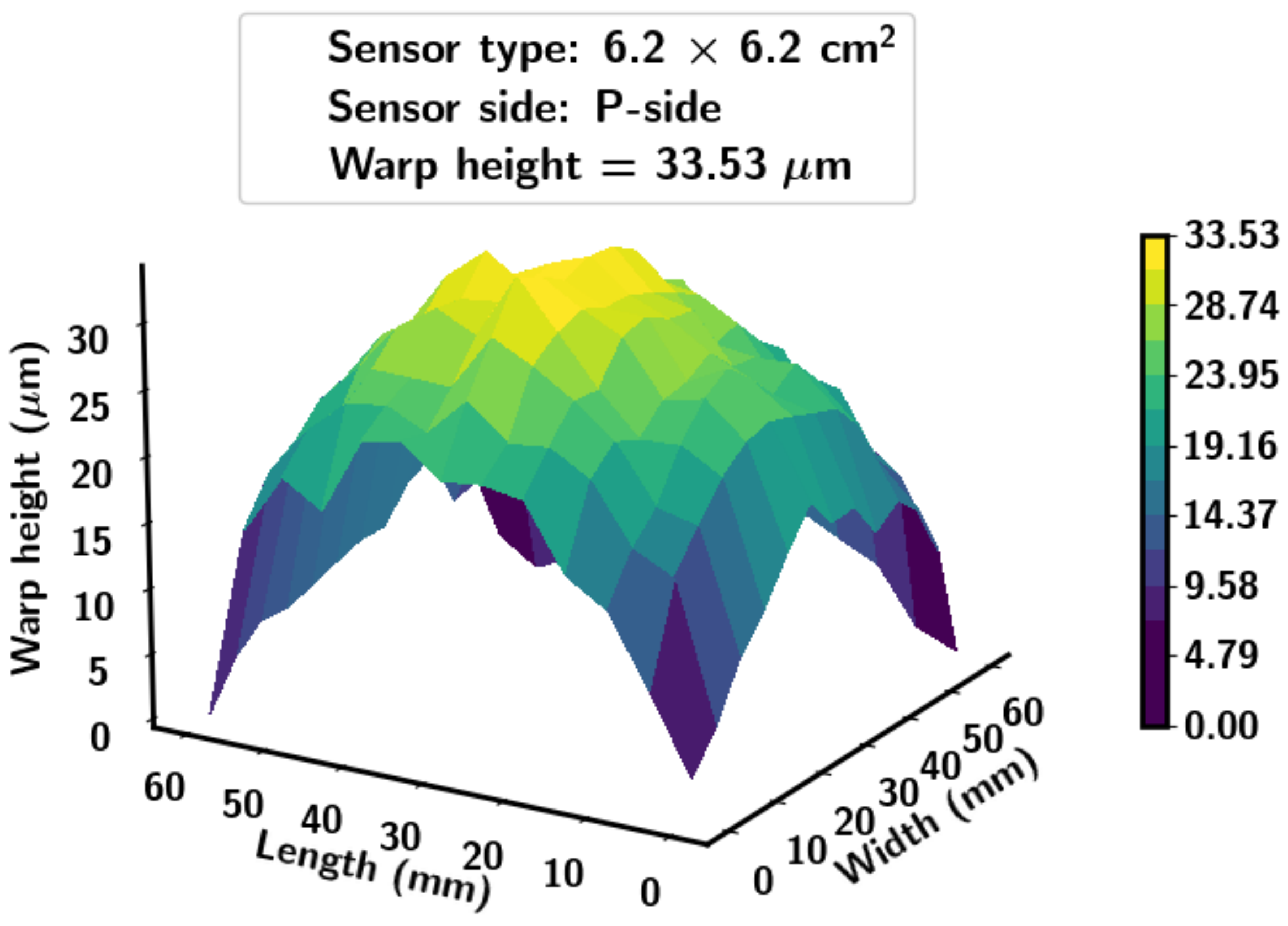}
\caption{Warp profile of p-side of sensor size 6.2 $\times$ 6.2 cm$^{2}$.  }
\label{fig:warp_data}
\end{figure}

The warp height is defined as the difference between the maximum and minimum distances from the reference plane surface. 
As can be seen in Figure \ref{fig:warp_data} the actual warp
profile of the p-side of one of the sensors variant 6.2 $\times$ 6.2 cm$^{2}$ and its warp height is obtained using the plane fit function.

Table \ref{tab:sensor_table} summarizes the warp height for a subset of sensors.
It has been observed that the mean value of maximum warp height for p-side of the sensor variants 6.2 $\times$ 2.2 cm$^{2}$, 6.2 $\times$ 4.2 cm$^{2}$ and 6.2 $\times$ 6.2 cm$^{2}$ is constrained 
below 50~$\mu$m. 
However, for the sensor variant 6.2 $\times$ 12.4 cm$^{2}$, the mean warp height was found slightly above 100 $\mu$m. 

\begin{table}[ht!]
\begin{adjustbox}{width=8.2cm, height = 2cm, center}
%\resizebox{\columnwidth}{!}{%
\begin{tabular}{l*{3}{c}r}\hline
& \multicolumn{1}{c}{}
& \multicolumn{1}{c}{\centering{Warp height ($\mu$m)}} \\
\hline
Sensor size (cm$^{2}$)& Amount of sensors & P-side & \\
\hline \hline \\[-1mm]
6.2 $\times$ 2.2& 35 & 23.0 $\pm$ 1.11 \\[3mm]
6.2 $\times$ 4.2& 55 & 35.2 $\pm$ 1.97\\[3mm]
6.2 $\times$ 6.2& 50 & 36.2 $\pm$ 1.94 \\[3mm]
6.2 $\times$ 12.4& 43 & 107.6 $\pm$ 5.21 & \\
\hline\\
\end{tabular}
\end{adjustbox}
\caption{Mean value of warp height for the different sensor variants.}
\label{tab:sensor_table}
\end{table}

The warp shape is the result of the mechanical stress induced on the silicon wafer during different manufacturing and polishing steps \cite{Ito, david}.

\section{Summary}
Optical inspection is one of the fundamental approaches to detect and study the silicon sensor's surface defects. The usage of Convolutional Deep Neural Networks for automated defect detection and classification yields precision results with an accuracy rate of 91.5$\%$ with an error of 8.5 $\%$. 

The study of defects present on the surface of the sensor allows us to sort them into different categories by introducing quality grade and quality score schemes.
The quality score was performed by weighting the different defect types. It was observed that with increasing sensor size, quality score is decreasing. The quality score is used to rank the sensors within the same quality grade to distribute the sensors among the tracking stations.

From the defective strips map of all sensor types, we have not observed any particular sensor area that has a higher number of defective strips. The defective strips are quite homogeneously distributed on the sensor's surface. However, clusters of consecutive defective strips have been found on the surface of all four sensor variants. From the distribution of the cluster analysis of consecutive defective strips, it has been found that most of the sensors from each sensor variant have clusters of less than 5 strips.

The 2D mapping of the defects was performed by converting the defect's pixel coordinate to its global geometrical coordinate. It has been observed 
that some of the defects appear in circular arrangement 
originating from a vacuum tool used to handle sensors during production at Hamamatsu Photonics.

There are as well further quality control activities taking place, e.g. electrical strip-by-strip characterization and module tests with radioactive source, which will demonstrate how much the defects observed by the optical inspection are affecting the sensor's overall performance. These investigations are currently ongoing and are yet to be concluded and published later in the future.

\section*{Acknowledgement}
We would like to thank the Detector Laboratory department at GSI for providing the necessary facility to carry out our work on sensor optical QA. We are in particular thankful to Dr. Christian Joachim Schmidt and Dr. Andrea Wilms for their support during this work.
From the CBM department, we are grateful to Dr. Olga Bertini and Marcel Bajdel for helping us with the operation of the inspection setup for
the sensor’s surface scanning. This project has received funding from the European Union’s Horizon 2020 research and innovation programme under grant agreement No. 871072  (CREMLINplus). The work was performed partly under the following grants BMBF 05P16VTFC1. We are also thankful to the GET-INvolved Programme for International Students and Researchers and Helmholtz Forschungsakademie Hessen f{\"u}r FAIR (HFHF), GSI Helmholtzzentrum f{\"u}r Schwerionenforschung, Frankfurt for providing financial support during the work.

%% Loading bibliography style file
%\bibliographystyle{model1-num-names}
%\bibliographystyle{cas-model2-names}

% Loading bibliography database
%\bibliography{cas-refs}

\end{document}